\documentclass[aps,pre,twocolumn,10pt,
 notitlepage,
 floatfix,
 nofootinbib,
  hidelinks,
  unicode=true,
 superscriptaddress]{revtex4-2}
\usepackage{float}
\usepackage{subfigure}
\usepackage{amssymb}
\usepackage{amsfonts}
\usepackage{amsmath}
\usepackage{amsthm}
\usepackage{epsfig}
\usepackage{float}
\usepackage{multirow}
\usepackage[usenames,dvipsnames]{color}
\usepackage{xr}
\usepackage{enumitem}
\usepackage{braket}
\usepackage{graphics,graphicx,xcolor,subfigure}
\usepackage{comment}
\usepackage{cancel}
\usepackage{orcidlink}

 \usepackage[]{hyperref}
 \hypersetup{colorlinks=true,
 	linkcolor=blue,
 	urlcolor=blue,
 	citecolor=blue,
 	pdfhighlight=/N
 }

\usepackage[normalem]{ulem}

\newcommand{\av}[1]{\langle{#1} \rangle}

\newcommand{\Ninf}{N_\mathrm{inf}}

\newcommand{\NSI}{N_\mathrm{SI}}
\newcommand{\kmax}{k_\text{max}}
\newcommand{\ximax}{\xi_\text{max}}
\newcommand{\kmin}{k_\text{min}}
\newcommand{\Ncore}{N_\text{c}}
\newcommand{\Pcore}{P_\text{c}}
\newcommand{\Nbridges}{N_\text{bridges}}
\newcommand{\avc}[1]{\langle{#1} \rangle_\text{c}}

\begin{document}

\title{Impacts of bridging nodes on the epidemic activation mechanisms}

\author{José Carlos M. Silva\orcidlink{0000-0002-9735-0108}}
\thanks{These authors contributed equally to this work.}
\affiliation{Departamento de F\'{\i}sica, Universidade Federal de Vi\c{c}osa, 36570-900 Vi\c{c}osa, Minas Gerais, Brazil}

\author{Diogo H. Silva\orcidlink{0000-0001-6639-6413}}
\thanks{These authors contributed equally to this work.}
\affiliation{Instituto de Ci\^{e}ncias Matem\'{a}ticas e de Computa\c{c}\~{a}o, Universidade de S\~{a}o Paulo, S\~{a}o Carlos, SP 13566-590, Brazil}
\thanks{\textit{Present address}: Instituto Federal Fluminense, Campus Quissamã, 28735-970,
		Piteiras-Quissamã, Rio de Janeiro, Brazil}

\author{Wesley Cota\orcidlink{0000-0002-8582-1531}}%
\affiliation{Departamento de F\'{\i}sica, Universidade Federal de Vi\c{c}osa, 36570-900 Vi\c{c}osa, Minas Gerais, Brazil}

\author{Francisco A. Rodrigues\orcidlink{0000-0002-0145-5571}}
\affiliation{Instituto de Ci\^{e}ncias Matem\'{a}ticas e de Computa\c{c}\~{a}o, Universidade de S\~{a}o Paulo, S\~{a}o Carlos, SP 13566-590, Brazil}

\author{Silvio C. Ferreira\orcidlink{0000-0001-7159-2769}}
\email{silviojr@ufv.br}
\thanks{Corresponding author.}
\affiliation{Departamento de F\'{\i}sica, Universidade Federal de Vi\c{c}osa, 36570-900 Vi\c{c}osa, Minas Gerais, Brazil}
\affiliation{National Institute of Science and Technology for Complex Systems, 22290-180, Rio de Janeiro, Brazil}
\thanks{{On leave at Instituto de Ci\^{e}ncias Matem\'{a}ticas e de Computa\c{c}\~{a}o, Universidade de S\~{a}o Paulo, S\~{a}o Carlos, SP 13566-590, Brazil}}

\begin{abstract}
Bridging nodes, which connect critical components of a network, play an important role in maintaining structural integrity and facilitating communication within the network, representing indirect yet relevant connections. Epidemic triggering mechanisms in networks often involve long-range mutual activation of hubs, mediated by paths composed of low-degree nodes. While low-degree nodes are abundant in networks, their role in bridging central nodes in epidemic activation mechanisms has not been thoroughly analyzed. Starting with a backbone network with a power-law degree distribution, we investigate the role of adding degree-2 bridging nodes that are preferentially attached to hubs. Our findings reveal that bridging nodes can mediate an indirect feedback interaction between hubs that modifies the epidemic localization  and activation mechanisms of the epidemic processes with recurrent infections. In particular, the collective activation  observed in the presence of waning immunity, which produces a finite epidemic threshold in power-law networks with degree exponent $\gamma>3$, is altered to a localized activation with a vanishing threshold. Our {numerical} results are {analytically} supported by the non-backtracking matrix properties that emerge in the recurrent dynamical message-passing theory.
\end{abstract}

\maketitle

\section{Introduction}
Spreading phenomena on networks and the underlying role played by contact heterogeneity have been extensively investigated~\cite{Pastor-Satorras2015} since the seminal work of Pastor-Satorras and Vespignani~\cite{Pastor-Satorras2001b}, a benchmark in network epidemiology that highlighted the foundational importance of hubs in driving epidemic activation. While much of the recent attention has focused on more complex structures, such as multilayer~\cite{DeDomenico2016}, temporal~\cite{Valdano2017}, and higher-order~\cite{Wang2024} networks, simple graphs with pairwise interactions still pose interesting and unanswered questions. For example, different types of epidemic localization~\cite{Goltsev2012,Silva2021,St-Onge2020a,Liu2019,Hindes2016} have been identified, in which a small fraction (a sub-extensive subset) of the network, composed of highly connected nodes, can sustain a global epidemic prevalence at finite values even in the limit of very low infection rates~\cite{Chatterjee2009,Boguna2013,Sander2016}.

Identification of influential spreaders for dynamical processes on networks is theoretically associated with network centralities~\cite{Lu2016} such as degree, eigenvector, betweenness, and others~\cite{DeArruda2018}. Epidemic localization is related to influential spreaders~\cite{Kitsak2010}, who are those individuals capable of promoting contagion much more than others. The number of contacts of a node, while important, is not necessarily the leading centrality measure to determine epidemic spreadability~\cite{DeArruda2014}. Kitsak et al.~\cite{Kitsak2010} presented several pieces of evidence and proposed that the $k$-core centrality is more suitable than degree centrality to identify the influential spreaders of the susceptible--infected--recovered (SIR) epidemic model in several synthetic and real networks. The $k$-core decomposition is obtained by iteratively removing nodes whose degree is smaller than $k$ until a connected subgraph in which every node has degree at least $k$ remains~\cite{Dorogovtsev2006}.
For random scale-free networks, the maximum $k$-core forms a densely connected subset and is associated with the influential spreaders~\cite{Kitsak2010}. Castellano and Pastor-Satorras~\cite{Castellano2012}, however, argued that, depending on the network structure, the susceptible--infected--susceptible (SIS) epidemic model is not led by a maximum $k$-core activation. Actually, for uncorrelated random networks with a power-law degree distribution with exponent $\gamma$, the locally activated hubs are the main actors in triggering the global epidemic processes for $\gamma > 5/2$, whereas the maximum $k$-core is responsible for $2 < \gamma < 5/2$~\cite{Castellano2012,Cota2018}.

Eigenvector centrality of the adjacency matrix is also a common proxy for node influence~\cite{Goltsev2012,Martin2014,Castellano2017,Silva2019} in the context of dynamical and, especially, epidemic processes on networks. However, it may overestimate the importance of hubs,
concentrating too much weight on them~\cite{Goltsev2012}.   Non-backtracking matrix eigenvector centrality softens this feature~\cite{Martin2014},  but it is also not free of overlocalization in real networks~\cite{PastorSatorras2020}.

The SIS dynamic activation is especially intricate for degree exponents $\gamma > 3$~\cite{Chatterjee2009,Boguna2013}. In this regime, sparsely distributed hubs and their neighbors form star (sub)graphs that can sustain epidemic activity for very long periods, exponentially long in the hub degree, permitting indirect (long-range) mutual infections through low-degree paths, thus granting an asymptotically null epidemic threshold even in the non--scale--free regime with $\gamma > 3$~\cite{Boguna2013}. {While the epidemic threshold is zero in the thermodynamical limit, an effective (size-dependent) epidemic threshold can be determined for finite-size systems~\cite{Ferreira2012,Boguna2013,Ferreira2016,Castellano2020}.} Models with immunity (permanent or not), such as SIR~\cite{Castellano2012} or susceptible--infected--recovered--susceptible (SIRS)~\cite{Ferreira2016,Silva2022}, exhibit local epidemic activity that is reduced when compared with SIS. Therefore, hubs are not able to mutually infect each other through long-range interactions, and a finite epidemic threshold with collectively driven (extensive component of the network) activation emerges for $\gamma>3$.

So far, the role of small-degree nodes seems to be secondary for epidemic activation. However, there is evidence that weak ties, links connecting loosely connected communities, play an important role in promoting propagation throughout the whole network~\cite{Zhao2010,Shu2012,Gallos2012} and in maintaining network integrity and communicability. Moreover, the importance of low-degree nodes has been acknowledged in different complex systems, such as the integration of brain activity networks~\cite{Ferraro2018} and transmission chains in COVID-19~\cite{Serafino2022}. On the one hand, using either degree or $k$-core centralities, nodes of degree 1 or 2 are the lowest ranked. On the other hand, nodes of degree 1 can increase the local epidemic lifetime of the nodes to which they are connected~\cite{Chatterjee2009,Friedrich2024}, while, more interestingly, degree-2 nodes can act as bridging nodes connecting hubs or highly active communities.

In this work, we investigate both numerically and analytically the localization and phase transitions of epidemic models with recurrent infections, namely SIS and SIRS, on complex networks. The networks start with a power-law degree distribution (the core network) and are subsequently augmented with an extensive number of degree-2 bridging nodes attached preferentially to hubs. Under linear attachment rules, these bridging nodes play no significant role. Moreover, the epidemic activation mechanism, initially governed by the maximum $k$-core, remains unchanged even after the addition of an extensive number of bridging nodes, regardless of the attachment rule. In contrast, when a superlinear attachment rule is applied, in which bridges are connected to nodes with probability proportional to their squared degrees, quantitatively and qualitatively relevant changes arise. In particular, in the SIRS dynamics, the collective activation observed in the original network is replaced by a localized activation centered on subgraphs composed of pairs of hubs sharing a common set of bridging nodes. Our findings are corroborated by mean-field theories.

The remainder of the paper is organized as follows. Section~\ref{sec:model} presents the model for building networks with bridging nodes added according to a hub-preferential attachment rule. The degree distribution is analytically derived for a core network with a power-law degree distribution. Section~\ref{sec:methods} presents the epidemic models, simulation methods, and metrics used to determine epidemic thresholds and localization patterns. Both simulations and mean-field theories are presented in Section~\ref{sec:results}. Finally, the discussions and concluding remarks are provided in Section~\ref{sec:conclu}.

\section{The network model}%
\label{sec:model}

\begin{figure}[th]
	\centering
	\includegraphics[width=0.99\linewidth]{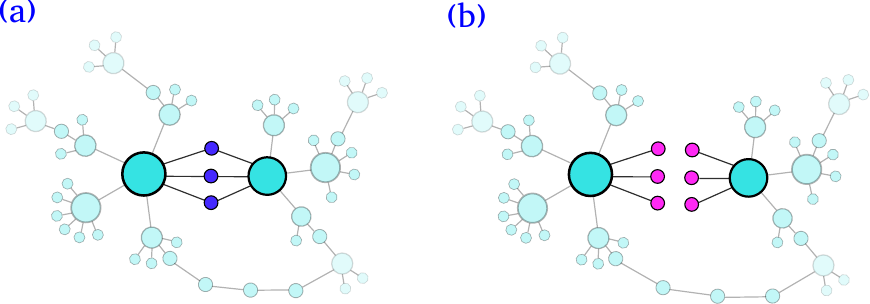}
	\caption{Schematic representation of the network model with bridges and {the concept of} drawbridges. (a) Bridging nodes of degree 2 (dark blue) are attached to the hubs of the core network (light blue). (b) {Every bridging node is replaced by two} nodes of degree 1, preserving the number of edges in the network{, forming a drawbridge}.}
	\label{fig:halfbridges}
\end{figure}

In order to investigate the role of bridging nodes in recurrent epidemic processes, we consider a model that starts with a \textit{core network} containing $\Ncore$ nodes, generated using the uncorrelated configuration model (UCM)~\cite{Catanzaro2005}, such that it initially does not exhibit degree correlations. The degree distribution of the initial core is $\Pcore(k)$, with lower and upper bounds $k_0$ and $k_\text{c}$, respectively. Next, $\Nbridges = f\Ncore$ bridging nodes of degree 2 are sequentially added, and each of their {unconnected stubs} is attached to {one of} the core nodes {chosen randomly} with probability {proportional to initial node degree, that is} $P_{\text{conn}} \propto k^{\nu}$, where $k$ is the {initial} degree of the selected core node. Figure~\ref{fig:halfbridges}(a) schematically shows a representation of the model.

Adding new connections can drastically change the degree distribution depending on the parameters $\nu$ and $\gamma$. Since only the original core is considered in the attachment rule, every new edge will connect to node $i$ of degree $k_i$ with a constant probability
\begin{equation}
	q_i=\frac{k_i^\nu}{\sum_{i=1}^{\Ncore} k_i^\nu} = \frac{k_i^\nu}{\Ncore\avc{k^\nu}}\,,
\end{equation}
where $\avc{\cdots}$ denotes the average over the core nodes. Since $q_i$ is constant, the probability that node $i$ receives $n_i$ new connections after the addition of $f\Ncore$ bridging nodes of degree 2 -- which correspond to $N_\text{e}=2f\Ncore$ new edges -- is therefore given by a binomial distribution:
\begin{equation}
	Q_i (n_i) = \binom{N_\text{e}}{n_i} q_i^{n_i}(1-q_i)^{N_\text{e}-n_i}\,,
\end{equation}
where we neglected the possibility that both edges of a bridging node attempt to connect to the same core node. Thus, the expected number of new edges added to node $i$ is given by
\begin{equation}
	\xi_i=\av{n_i} = \frac{2f k_i^\nu}{\avc{k^\nu}},.
	\label{eq:kiprime}
\end{equation}
Using a continuous approach where the degree distribution of the core is $\Pcore(k)\simeq(\gamma-1) k_0^{\gamma-1} k^{-\gamma}$, we have
\begin{equation}
	\avc{k^\nu} =
	\left\{
	\begin{array}{ccc}
	\dfrac{(\gamma-1)k_0^\nu}{\nu+1-\gamma} \left(\dfrac{\kmax}{k_0}\right)^{\nu+1-\gamma} &, & \gamma< \nu+1 \\
	& & \\
    \dfrac{(\gamma-1)k_0^\nu}{\gamma-\nu-1}  &,& \gamma> \nu +1
	\end{array}\,,
	\right.
	\label{eq:kinumed}
\end{equation}
{where $\kmax$ is the average degree of the most connected node in the original core.}%
\footnote{{Note that, due to the finite size of the networks, an upper cutoff for the average maximum degree, $\kmax \ll \Ncore$, will necessarily occur, even if $k_\text{c} \gg \Ncore$. This cutoff depends on the core degree distribution and its  cutoff~\cite{barabasi2016network,Silva2019}.}}

The added edges distribution is given by
\begin{equation}
	\psi(\xi) =  \Pcore(k) \left|\frac{dk}{d\xi}\right| =
	\frac{(\gamma-1)}{\nu} \xi_0^{\frac{\gamma-1}{\nu}}\xi^{-\omega};\qquad \xi\ge \xi_0\,,
\end{equation}
where  the power-law exponent is given by
\begin{equation}
	\omega=\frac{\gamma+\nu-1}{\nu}
	\label{eq:omega}
\end{equation}
and the lower bound of $\xi$ is given by
\begin{equation}
	\xi_0 = \dfrac{2fk_0^\nu}{\avc{k^\nu}}
		= \left\{
	\begin{array}{ccc}
		2f\dfrac{\nu+1-\gamma}{\gamma-1} \left(\dfrac{k_0}{\kmax}\right)^{\nu+1-\gamma} &, & \gamma< \nu+1 \\
		& & \\
		2f\dfrac{\gamma-\nu-1}{\gamma-1}  &,& \gamma> \nu +1
	\end{array}.
	\right.
\end{equation}
{The largest degree increment is given by Eq.~\eqref{eq:kiprime} evaluated at $k_i=\kmax$, then given by
	\begin{equation}
		\xi_\text{max}
		= \frac{2 f\, \kmax^\nu}{\avc{k^\nu}}
		\sim
		\left\{
		\begin{array}{ccc}
			\kmax^{\gamma-1} & , & \gamma < \nu + 1 \\
			\kmax^{\nu}      & , & \gamma > \nu + 1
		\end{array}
		\right.
		\label{eq:ximax}
	\end{equation}
	{Since $\gamma>2$}, in both regimes we asymptotically have $\xi_\text{max}\gg \kmax$ in the superlinear regime with $\nu>1$, implying that total degree of hubs is governed by the newly added connections such that $k'_\text{max}=\kmax+\xi_\text{max} \simeq \xi_\text{max}$.}

The new degree of core nodes, $k' = k + \xi$, is therefore the convolution of two power-law distributions whose tail is governed by the exponent $\gamma' = \min(\gamma, \omega)$, such that
\begin{equation}
	\gamma'=
	\left\{
		\begin{array}{ccc}
			\gamma &, & \nu\le 1 \\
			& & \\
			\dfrac{\gamma+\nu-1}{\nu} &,& \nu > 1
		\end{array}.
		\right.
		\label{eq:gammalinha}
	\end{equation}
In particular, a scale-free network with $2 < \gamma' \le 3$ is produced when $\nu + 1 < \gamma < 2\nu + 1$, while {$\gamma'<2$,  representing} a diverging average degree distribution {in the thermodynamical limit when both $\Ncore$ and $k_\text{c}$ diverge,}  occurs  if $\omega < 2$, which means $\gamma < \nu + 1$. In the latter case, a \textit{winner-take-all}~\cite{barabasi2016network} mechanism {is at play}, where a few nodes absorb most of the new connections, forming superhubs with degrees of the same order as the network size.

Finally, for $\nu>1$, the crossover to the regime $\Pcore'(k')\sim k'^{-\gamma'}$ will happen when $\xi^* \simeq k^*$ that leads to
\begin{equation}
	k^*=
	\left\{
	\begin{array}{ccc}
		C   &, & \gamma>\nu+ 1 \\
		& & \\
		C  \left[\dfrac{\kmax}{k_0}\right]^{\theta} &,& \gamma <\nu+1
	\end{array}\,,
	\right.
	\label{eq:kstar}
\end{equation}
where $C$ is a constant given by
\[
C= \left[\dfrac{(\gamma-1)k_0^{\nu}}{2f(\gamma-\nu-1)}\right]^{\frac{1}{\nu-1}}\,,
\]
and  the exponent $\theta$ is given by
\begin{equation}
	\theta = 1-\frac{\gamma-2}{\nu-1} <1\,.
\end{equation}

\begin{figure}[hbt!]
	\includegraphics[width=0.99\linewidth]{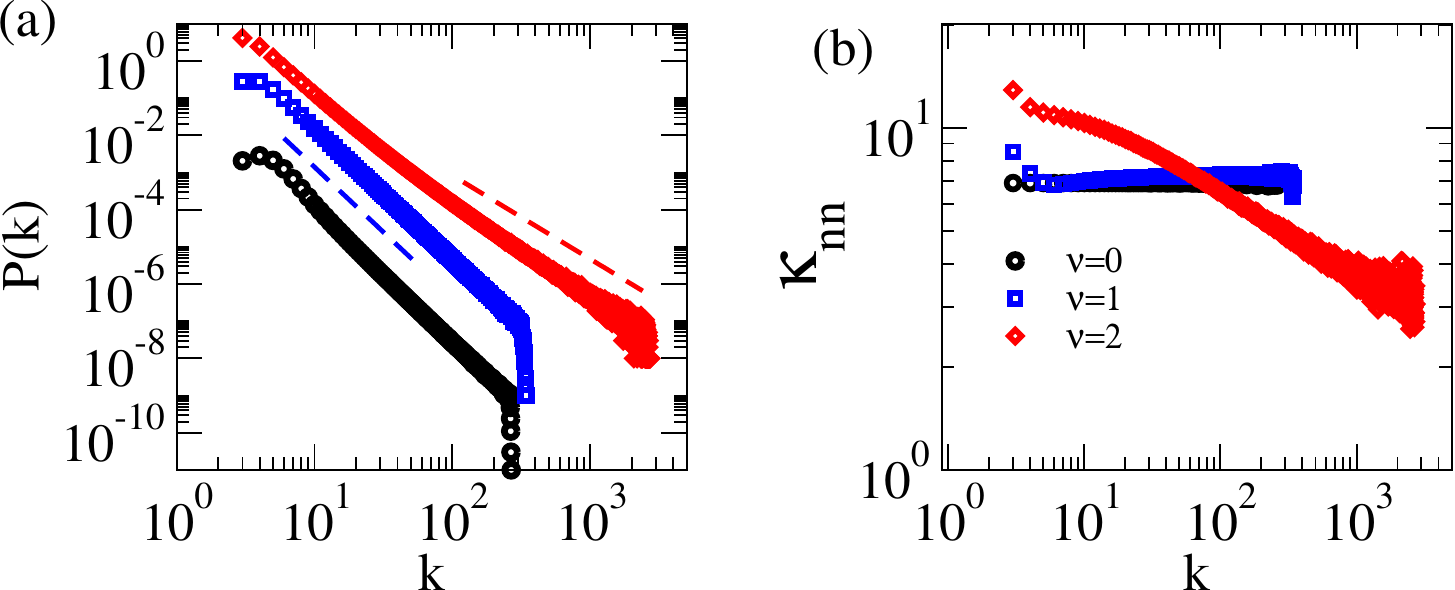}
	\caption{(a) Degree distribution and (b) degree correlation analysis for a core network doped with a fraction $f=0.5$ of bridging nodes. The original core uses a UCM network with $\Ncore=10^6$ nodes, minimum degree $k_0=3$, degree exponent $\gamma=3.5$, and a rigid upper cutoff such that $\kmax \sim N_\text{c}^{1/\gamma}$. Random attachment of bridging nodes ($\nu=0$) is compared with linear ($\nu=1$) and super-linear ($\nu=2$) preferential attachment rules. Dashed lines represent power-law decays with exponents $\gamma'=3.5$ and $\gamma'=2.25$, corresponding to the predictions for $\nu=1$ and $\nu=2$, respectively. Degree distributions were shifted to improve visibility.}	\label{fig:NET}
\end{figure}

The degree distribution of the modified network model is shown in Fig.~\ref{fig:NET}(a) for a UCM core with a power-law degree distribution $P_\text{c}(k)\sim k^{-\gamma}$, exponent $\gamma=3.5$, and a rigid cutoff given by $\Ncore P_\text{c}(k_\text{c}) = \text{const.}$, which leads to a core without degree outliers and a maximum degree $\kmax\sim \Ncore^{1/\gamma}$~\cite{Silva2019}. This choice avoids the presence of outliers in the core's degree distribution and makes the analysis of the epidemic threshold unambiguous for this range of degree exponents~\cite{Silva2019}. A linear preferential attachment ($\nu=1$) yields degree distributions whose tails scale as in random attachment, being equivalent to the original UCM core, while for a super-linear preferential attachment ($\nu=2$), the degree distribution exhibits a heavier tail, in agreement with the analytical scaling given by Eq.~\eqref{eq:gammalinha}.

We also analyzed network correlations by investigating the average degree of nearest neighbors, $\kappa_\text{nn}(k)$, as a function of node degree~\cite{Pastor_Satorras2001}. For disassortative or assortative correlation patterns, $\kappa_\text{nn}(k)$ is a decreasing or increasing function of the degree, respectively, while it remains constant for uncorrelated (neutral) networks. The linear attachment preserves the uncorrelated pattern of the original UCM core, whereas the super-linear case, where the degree distribution becomes scale-free, introduces a strong disassortative regime.

\section{Methods}
\label{sec:methods}

We investigate the SIRS dynamics, in which infected nodes recover spontaneously with rate $\mu$, while recovered nodes spontaneously become susceptible with rate $\alpha$. Susceptible nodes are infected at rate $\lambda$ per infectious contact. The limit $\alpha \rightarrow \infty$ yields the SIS dynamics, characterized by instantaneous loss of immunity after recovery. This recurrent infection dynamics presents an absorbing state in which the disease is eradicated. Since the system is finite, the simulations always fall into this state after sufficiently long times. To avoid this issue, we adopted the hub reactivation method~\cite{Sander2016}, in which the most connected node in the network is reinfected when the dynamics enters the absorbing state. We considered the most connected node of the original core to be reactivated. This strategy leads to a steady regime after a sufficiently long relaxation time $t_\text{rlx}$, known as the quasi-stationary (QS) state~\cite{Sander2016}. Averages are computed over an averaging time $t_\text{av}$ following the relaxation time $t_\text{rlx}$. Typical values of $t_\text{rlx}$ and $t_\text{av}$ used in this work are $10^6$ and $10^7$, respectively. This setup applies to the range of $\lambda$ values in which long-lasting activity emerges. Higher relaxation and averaging times are typically used for lower values of $\lambda$, while shorter averaging times are sufficient for higher values of $\lambda$, due to the shorter inter-event time. The computer implementation of the stochastic simulations is presented in Appendix~\ref{app:simu}.

The focus of this work is to investigate how bridging nodes alter epidemic localization and the activation mechanisms near the epidemic threshold. To perform a comparative analysis, all QS quantities are computed only for the nodes belonging to the original core, i.e., excluding the bridging nodes. We analyzed the dynamical susceptibility defined as~\cite{Sander2016}
\begin{equation}
	\label{eq:Susceptibility}
	\chi = \Ncore\frac{\av{\rho^2}-\av{\rho}^2}{\av{\rho}}\,,
\end{equation}
where  $\av{\rho^n}$ is the order parameter moments computed in the QS state.The curves of susceptibility versus infection rate exhibit a peak, which is used as an estimate of the epidemic threshold $\lambda_\text{c}$~\cite{Silva2022,Cota2017}.

The epidemic localization can be investigated by considering the activity vector $\boldsymbol{\rho}=(\rho_1, \rho_2, \ldots, \rho_{\Ncore})$~\cite{Silva2021}, defined as
\begin{equation}
	\label{eq:rho}
	\rho_i = \frac{1}{t_\text{av}}\int_{t_\text{rlx}}^{t_\text{rlx}+t_\text{av}}\sigma_i(t)dt\,,
\end{equation}
where $\sigma_i(t) = 1$ if vertex $i$ is infected at time $t$, and $0$ otherwise. The normalized activity vector (NAV) $\boldsymbol{\phi}$ is given by
\begin{equation}\label{eq:nav}
	\phi_i = \frac{\rho_i}{\sqrt{\sum_j \rho_j^2}}\,.
\end{equation}
We investigate the epidemic localization considering the inverse participation ration (IPR) defined as~\cite{Goltsev2012}
\begin{equation}\label{eq:IPR_NAV}
	Y_4 = \sum_{i=1}^{\Ncore} \phi_i^4\,.
\end{equation}
For a delocalized vector, one has $Y_4 \sim \Ncore^{-1}$, while for a vector localized on a finite number of components, one has $Y_4 \sim \text{const.}$~\cite{Goltsev2012}. It is also possible for a vector to be localized on a sub-extensive component, for which $Y_4 \sim \Ncore^{-a}$ with $0 < a < 1$~\cite{Castellano2012}.

\section{Results}
\label{sec:results}

\subsection{Stochastic simulations}
\label{subsec:simu}

\begin{figure}[hbt]
	\includegraphics[width=0.99\linewidth]{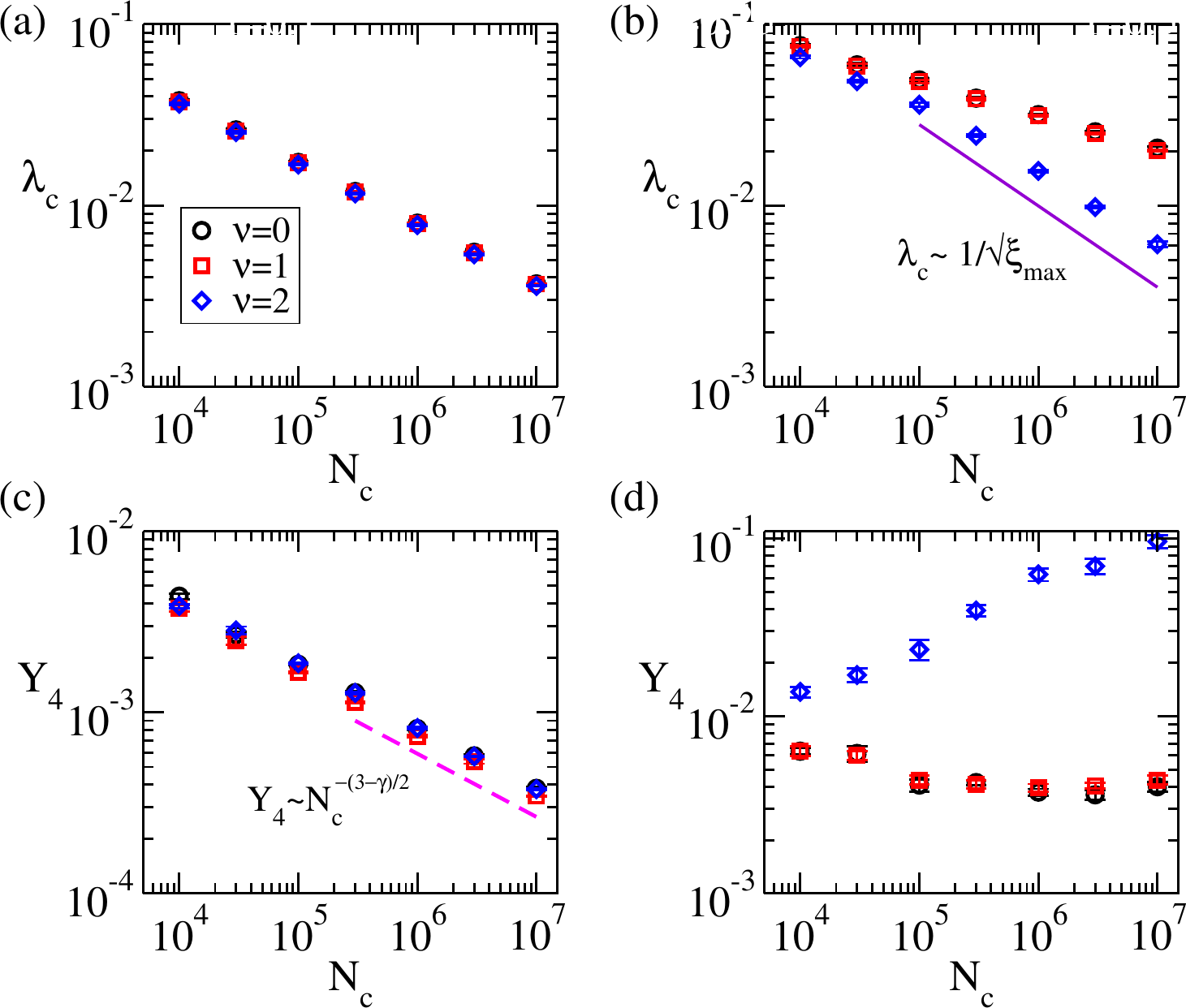}
	\caption{(a,b) Epidemic threshold and (c,d) IPR as functions of the core size for different attachment rules in the SIS dynamics. Random ($\nu=0$), linear ($\nu=1$), and super-linear ($\nu=2$) attachments are shown. The fraction of bridging nodes is $f=0.25$, and the core is a UCM network with (a,c) $\gamma=2.3$ or (b,d) $\gamma=2.8$, for $\kmin=3$ and $k_\text{c}=2\sqrt{\Ncore}$. The line in (b) represents the decay determined by the activation of the largest hub of degree $\ximax$, while in (c) it corresponds to the scaling $Y_4\sim \Ncore^{-(3-\gamma)/2}$, compatible with localization on the maximum $k$-core.}

	\label{fig:compara_nus}
\end{figure}

Different attachment rules are investigated for the SIS dynamics, as shown in Fig.~\ref{fig:compara_nus}, where random ($\nu=0$) and linear ($\nu=1$) rules are compared with the super-linear ($\nu=2$) attachment rule, using different degree exponents for the original core. For $\gamma=2.3$, where the epidemic activation is driven by the maximum $k$-core component of the original core, the addition of bridges does not change either the epidemic threshold or the activity localization measured by the IPR. Since the hubs that compose the maximum $k$-core are activated by direct interactions, the next-nearest-neighbor interactions mediated by bridges are irrelevant. {In this activation regime ones has $\lambda_\text{c}\sim 1/\av{k^2}\sim \kmax^{-(3-\gamma)} \sim \Ncore^{-(3-\gamma)/2} \sim \Ncore^{-0.35}$ for the structural cutoff $\kmax\sim \sqrt{\Ncore}$, consistent with the numerical scaling exponent estimated for the last 3 point of Fig.~\ref{fig:compara_nus}(a) that is approximately $\lambda_\text{c}\sim \Ncore^{-0.34}$. }

For $\gamma=2.8$, where the activation of the original core is triggered by weakly interacting hubs~\cite{Boguna2013,Castellano2020}, the random and linear attachments are equivalent, showing no significant differences with respect to the original UCM core. However, the super-linear attachment ($\nu=2$) leads to more localized (higher IPR) epidemics, and a different scaling of the epidemic threshold emerges. Indeed, the loosely interacting hubs produce an epidemic threshold similar to that of a star graph, consisting of a center connected to $\ximax$ neighbors of degree 1, which scales as $\lambda_\text{c}\sim 1/\sqrt{\ximax}$~\cite{Castellano2012}. The scaling given by Eq.~\eqref{eq:ximax} is $\ximax\sim \kmax^{\gamma-1} \sim \Ncore^{(\gamma-1)/2}\sim \Ncore^{0.9}$ for $\gamma=2.8$ and a structural cutoff $\kmax\sim \sqrt{\Ncore}$.  So, the threshold scaling is asymptotically $\lambda_\text{c}\sim \Ncore^{-0.45}$, sharply distinct from the maximum $k$-core activation regime where $\lambda_\text{c}\sim \Ncore^{-(3-\gamma)/2}\sim \Ncore^{-0.1}$ and consistent with the numerical estimate for the last 3 points of Fig.~\ref{fig:compara_nus}(b)  given by $\lambda_\text{c}\sim \Ncore^{-0.41}$.  Physically, this implies that both hubs are activated almost independently due to a feedback interaction, in which the hub infects its neighbors, which in turn reinfect the hub, as discussed in Refs.~\cite{Boguna2013,Ferreira2016}, rather than through the mutual interaction mediated by the bridges. This prediction is confirmed in Fig.~\ref{fig:compara_nus}(b), where the scaling of the epidemic threshold follows very closely that of the corresponding star graph.

\begin{figure}[hbt]
	\includegraphics[width=0.99\linewidth]{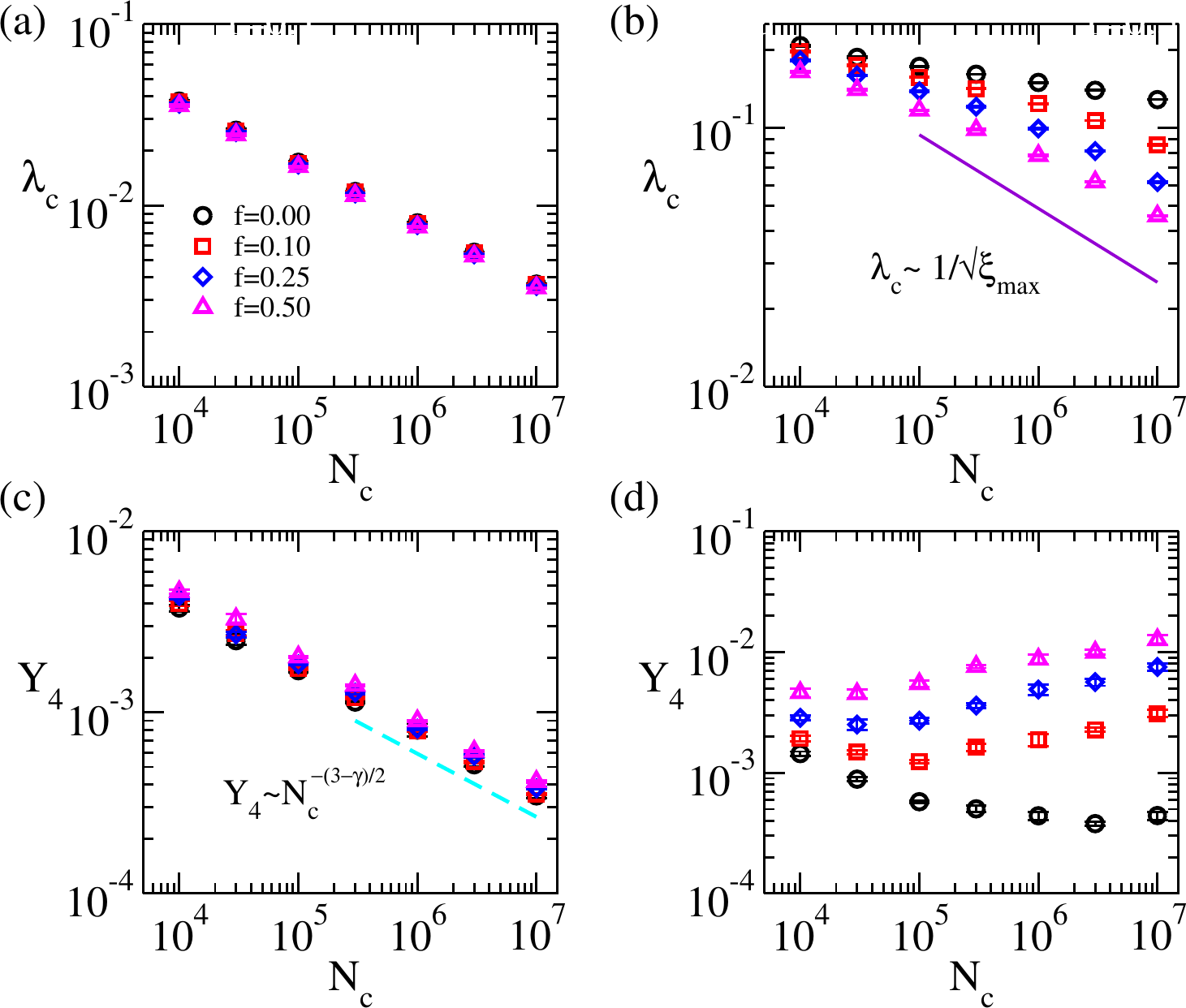}
	\caption{Epidemic threshold (top) and IPR of the NAV (bottom) as functions of the core size $\Ncore$ for the SIS dynamics, considering different fractions $f$ of added bridging nodes. A super-linear attachment with $\nu=2$ is used. Left panels: the core is a UCM network with $\gamma=2.3$, $k_0=3$, and $k_\text{c}=2\sqrt{\Ncore}$. Right panels: the core is a UCM network with $\gamma=3.5$, $k_0=3$, and $\kmax\sim \Ncore^{1/\gamma}$. Lines have the same meaning as in Fig.~\ref{fig:compara_nus}.}
	\label{fig:compara_fs}
\end{figure}

The impact of increasing the number of bridges in the SIS model is shown in Fig.~\ref{fig:compara_fs} for distinct activation regimes. For the case of maximum $k$-core activation, represented by $\gamma=2.3$, the effect of adding bridges is minor, leading to essentially the same epidemic threshold and localization patterns as those of the original UCM core, corroborating the robustness of maximum $k$-core activation in the SIS dynamics. Conversely, for the case of long-range activation of hubs, investigated for $\gamma=3.5$, the localization increases while the epidemic threshold decreases as more bridging nodes are included. This occurs due to the increased degree of the most connected node in the network, whereas the indirect interactions mediated by bridges have a negligible effect (see discussion on drawbridges below). As in Fig.~\ref{fig:compara_nus}(b), the epidemic threshold is consistent with the activation of the most connected node, whose degree, according to Eq.~\eqref{eq:ximax}, yields
\[
\lambda_\text{c}\sim \frac{1}{\sqrt{\ximax}} \sim \kmax^{-\nu/2} \sim \Ncore^{-\nu/2\gamma} = \Ncore^{-2/7}\,,
\]
for $\gamma=3.5$ and a rigid cutoff scaling as $\kmax\sim \Ncore^{1/\gamma}$.


\begin{figure}[hbt]
	\includegraphics[width=0.9952\linewidth]{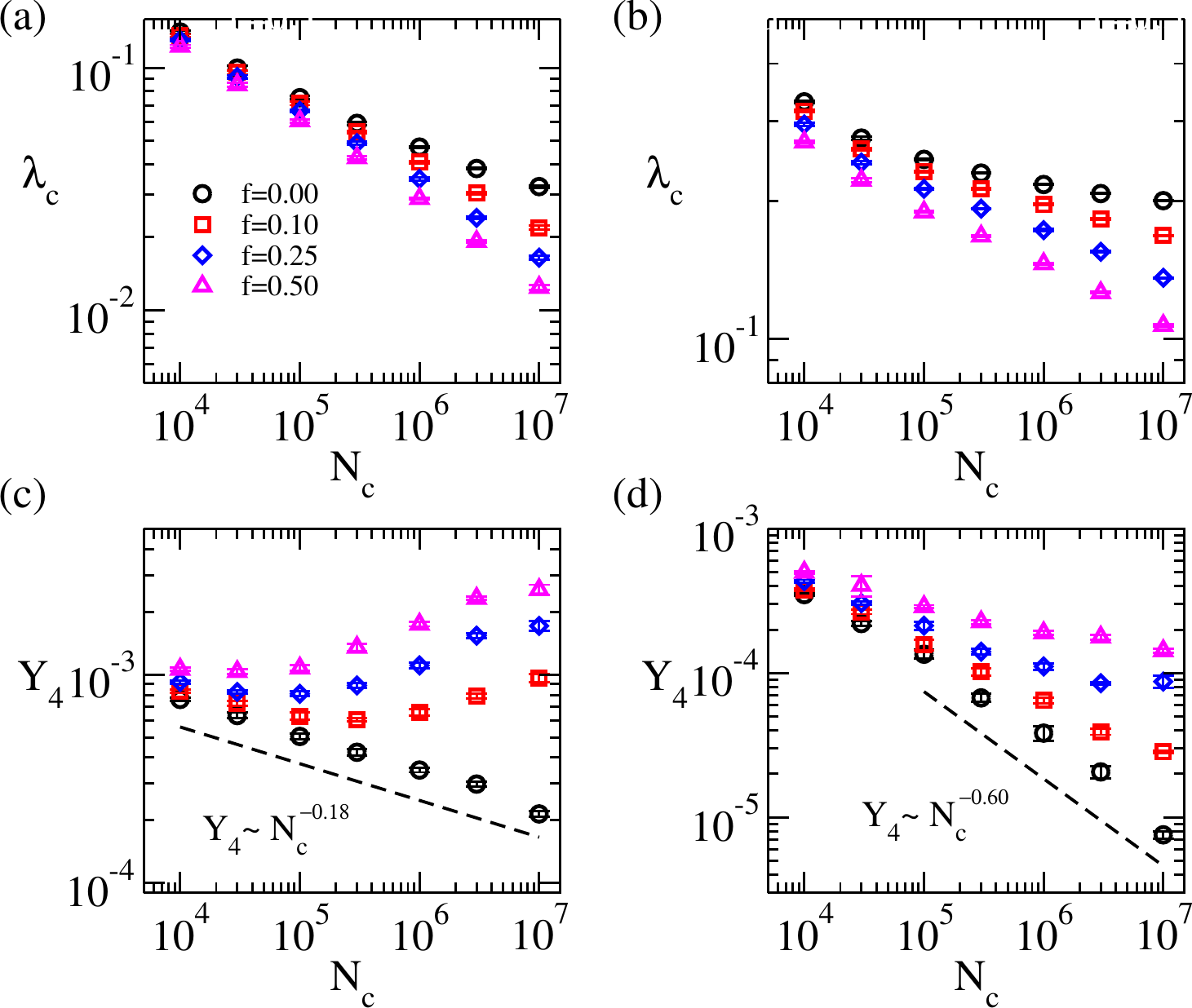}
	\caption{Epidemic threshold (top) and IPR of the NAV (bottom) as functions of the core size $\Ncore$ for SIRS dynamics with $\alpha=0.1$ and a super-linear attachment rule with $\nu=2$. The core is a UCM network with $k_0=3$ and degree exponent $\gamma=2.8$ with $k_\text{c}=2\sqrt{\Ncore}$ (left) or $\gamma=3.5$ with $\kmax\sim \Ncore^{1/\gamma}$ (right). Dashed lines represent the scaling exponents obtained from a power-law regression of the IPR.}
	\label{fig:sirs}
\end{figure}

Bridging nodes play a particularly significant role in the presence of waning immunity. We analyzed the SIRS dynamics with a waning immunity rate $\alpha<1$, for which the differences between the SIS and SIRS models on UCM networks become noticeable~\cite{Sander2016,Silva2022}. The conclusions remain similar to those for the SIS case when $\gamma<2.5$, as the activation is still driven by the maximum $k$-core mechanism~\cite{Sander2016}. For larger values of $\gamma$, however, the outcomes differ substantially. Figure~\ref{fig:sirs} shows the epidemic threshold and localization analysis for $\gamma>5/2$, covering both the scale-free and non–scale-free regimes, with $\gamma=2.8$ and $\gamma=3.5$, respectively.

For $\gamma=2.8$, shown in Figs.~\ref{fig:sirs}(a,c), the original core dynamics exhibit subextensive localization, with the IPR scaling approximately as $Y_4 \sim \Ncore^{-0.18}$. This pattern is modified by the presence of bridging nodes, leading to localization on a finite number of nodes, with $Y_4$ approaching an asymptotically finite value. The threshold scaling decreases more rapidly compared with the dynamics without bridging nodes. This effect becomes even more pronounced for $\gamma=3.5$: while activation in the original core is collective, yielding an asymptotically finite epidemic threshold~\cite{Sander2016,Silva2022}, bridging nodes lead to an asymptotically vanishing threshold, with localized activity converging to a finite asymptotic IPR. {While bridging suggest reduction of the localization with respect to the original core, as bridges are added, the hub degree simultaneously increases, due to the attachment towards high-degree nodes of the original core. This superlinear attachment rule promotes the emergence of degree outliers, which enhance epidemic localization  in SIS dynamic on the most connected node of the emerging, bridged networks. For SIRS model, where isolated hubs are not able to sustain long-term (exponential in degree) local activity around hubs~\cite{Ferreira2016}, mutual reinfection between hubs mediated by bridges is bursted, leading to strong localization around bridged hubs.}

These outcomes for the SIRS model  can be rationalized as follows. The epidemic lifespan of an isolated hub (a star graph) increases algebraically with the number of neighbors, $\tau \sim k^{\alpha/\mu}$; see Ref.~\cite{Sander2016} for details. In the original core, the hubs are sufficiently separated to block long-range interactions at very small infection rates. Bridging nodes, however, alter this scenario through a threefold effect: (i) they increase the epidemic lifespan of hubs by effectively raising their degree; (ii) they reduce the distance between hubs to $\ell=2$; and (iii) the presence of multiple bridging nodes amplifies the interaction strength, making it more significant than that of a single direct interaction. The latter two effects are the dominant factors, as discussed below. Consequently, a feedback mechanism arises in which mutual interactions between hubs are mediated by bridging nodes.

\begin{figure}[hbt]
	\includegraphics[width=0.95\linewidth]{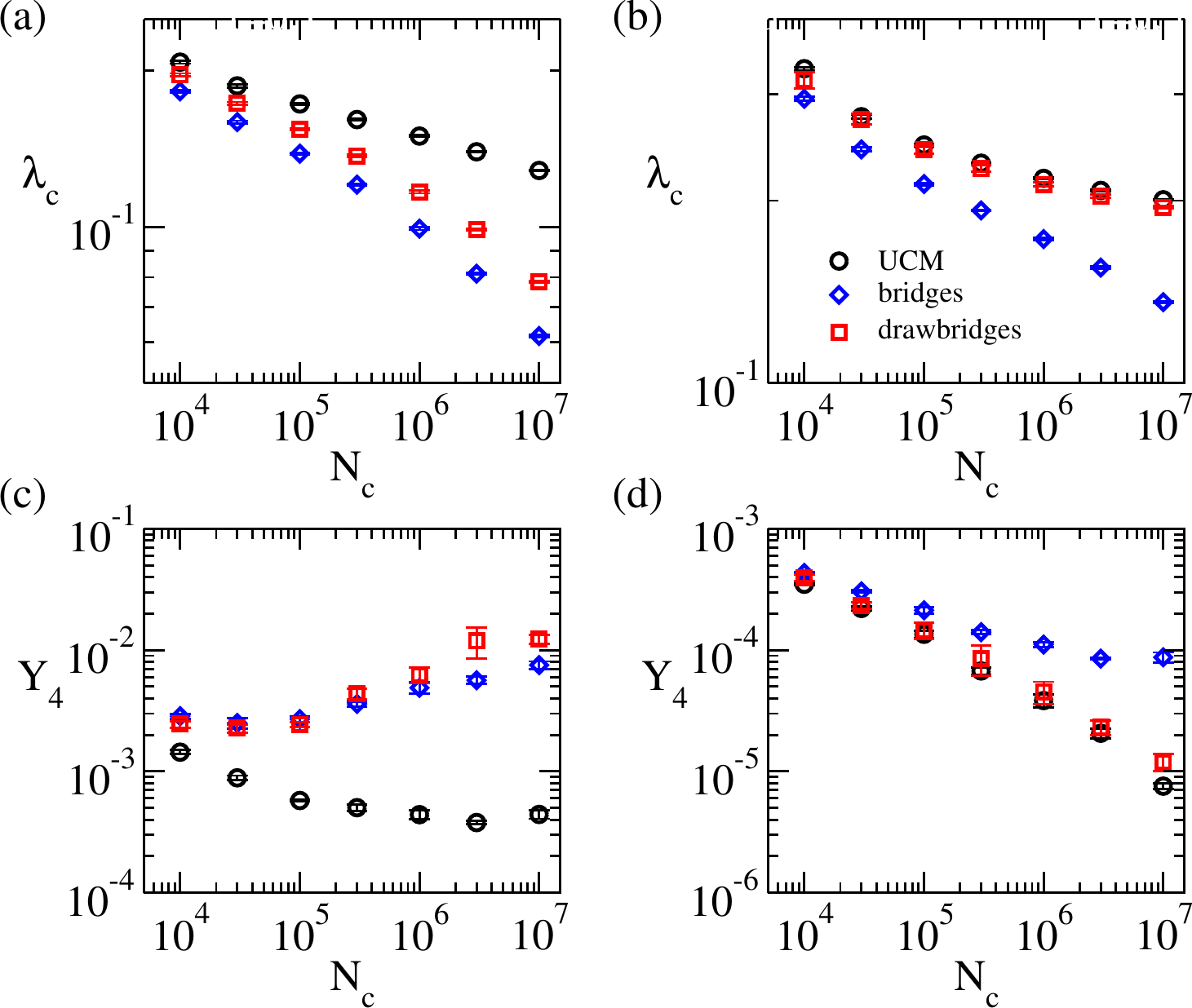}
	\caption{Comparison between bridging nodes and drawbridges for (a,c) SIS and (b,d) SIRS dynamics with $\alpha=0.1$; see Fig.~\ref{fig:halfbridges}(b). Epidemic threshold (top) and IPR of the NAV (bottom) as functions of the core size $\Ncore$ are presented. A super-linear attachment rule with $\nu=2$ is adopted for a fraction $f=0.25$ of added bridging nodes. The core is a UCM network with $\gamma=3.5$, $k_0=3$, and $\kmax\sim \Ncore^{1/\gamma}$.}
	\label{fig:SIRShalfbridges}
\end{figure}

In order to explicitly resolve the roles played by the increase in the hub's degree and the feedback mutual interaction between hubs mediated by bridging nodes, we construct networks where the bridging nodes of degree 2 are transformed into two nodes of degree 1, preserving the number of edges and the degree sequence of the network, as schematically shown in Fig.~\ref{fig:halfbridges}(b). These new configurations are called \textit{drawbridges}, for short. Figure~\ref{fig:SIRShalfbridges} compares the effects of bridging nodes and drawbridges on SIS and SIRS dynamics for an original core with degree exponent $\gamma=3.5$. For SIS dynamics, one observes a reduction of the epidemic threshold and an increase in the activity localization, measured by the IPR of the NAV. The bridging nodes and drawbridges present very similar outcomes for both threshold and IPR scaling. Indeed, SIS dynamics in this regime are governed by localized hub activity, and bridging plays only a minor role. For SIRS dynamics, however, the bridging effect is strong. Drawbridges alter very little the IPR and yield essentially the same threshold as the original UCM core, while the bridging nodes completely change the picture, as discussed previously. The conclusion is that the alteration observed for SIRS in Fig.~\ref{fig:sirs} is governed by a feedback interaction between hubs mediated by the bridges, forming an analogue of a maximum $k$-core where the hubs are two edges apart.

\subsection{Mean-field analysis}
\label{subsec:mf}

From the dynamical point of view, the spectral properties of matrices associated with connectivity are directly related to mean-field theories~\cite{Pastor-Satorras2015}. Two properties are especially relevant: the principal eigenvector (PEV) and the corresponding largest eigenvalue (LEV)~\cite{Goltsev2012,Castellano2017}, associated with the Jacobian matrices obtained during linear stability analysis. The adjacency matrix $A_{ij}$, defined as $A_{ij}=1$ if nodes $i$ and $j$ are connected and $A_{ij}=0$ otherwise, appears in the transition analysis of the quenched mean-field theory (QMF) for SIS~\cite{Goltsev2012} and SIRS~\cite{Silva2022}. According to the QMF theory, the epidemic threshold is given by the inverse of the LEV, while the epidemic prevalence of node $i$ near the transition is proportional to the corresponding PEV component~\cite{Goltsev2012,VanMieghem2012_b,Silva2019,Silva2022}.

The QMF theories may not be suitable for SIRS dynamics with $\alpha<1$~\cite{Sander2016}. Actually, SIRS dynamics on power-law networks are well described by the recurrent dynamical message passing (rDMP) theory~\cite{Silva2022}, where the Jacobian is related to the {non-backtracking} matrix~\cite{Hashimoto1989,Munik2015}, given by
\begin{equation}
	B_{j\rightarrow i, k\rightarrow l}=\delta_{jl}(1-\delta_{ik})\,,
	\label{eq:Hashimoto}
\end{equation}
where $j\rightarrow i$ means that there is an edge pointing from node $j$ to node $i$. {The non-backtracking matrix  is usually nonsymmetric, has dimension $N\av{k}$, and the condition $i=k$
	enforces continuity of a walk on it while  $j\ne l$ forbids immediate backtracking~\cite{Munik2015}.} The epidemic threshold is given by the inverse of the LEV~\cite{Munik2015}, while the epidemic prevalence is proportional to the corresponding component of the PEV of the {non-backtracking} matrix~\cite{Silva2022}.

\begin{figure}[hbt]
	\centering
	\includegraphics[width=0.99\linewidth]{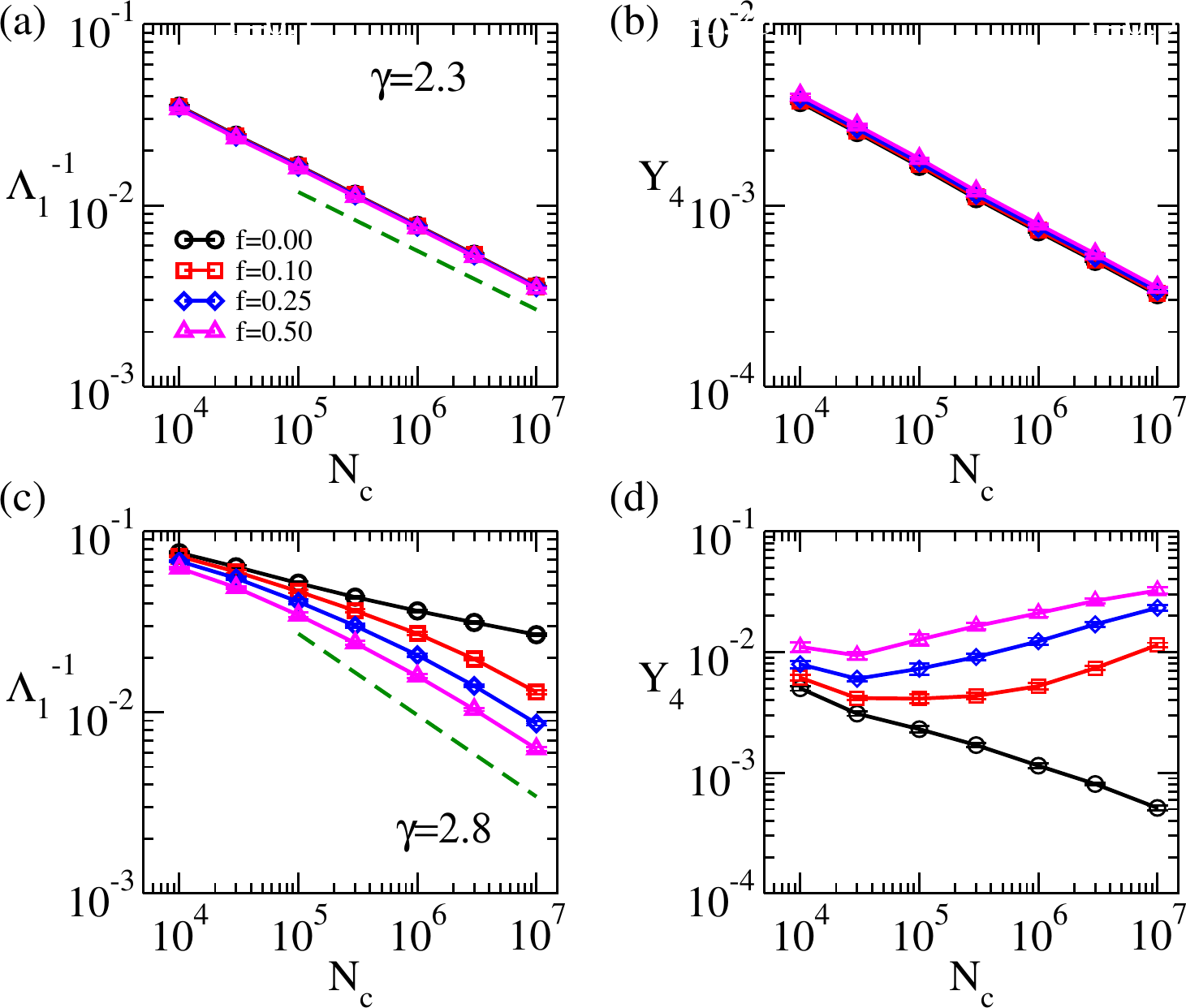}\\
	\includegraphics[width=0.99\linewidth]{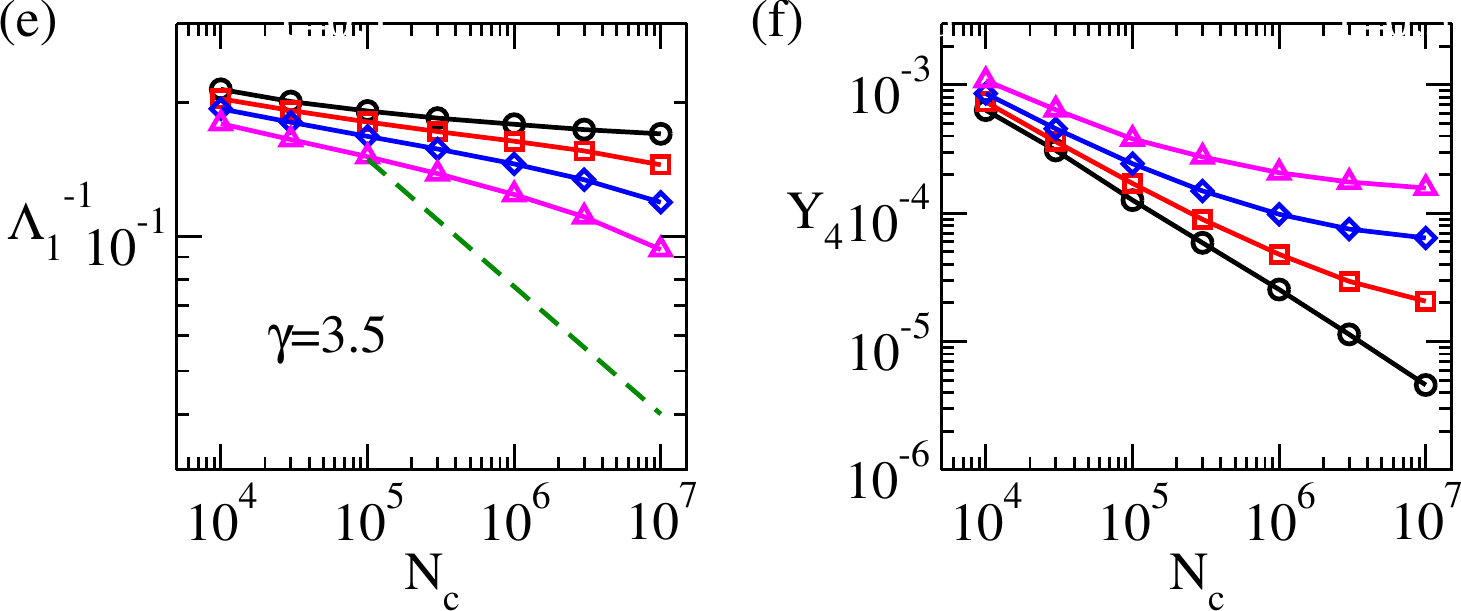}
	\caption{Spectral analysis of {non-backtracking} matrices for different amounts of bridging nodes, considering power-law networks with degree exponent (a,b) $\gamma=2.3$, (c,d) $\gamma=2.8$, and (e,f) $\gamma=3.5$. The inverse of the LEV $\Lambda^{-1}$ (top) and the IPR of the PEV $Y_4$ (bottom) as functions of the core size $\Ncore$ are shown for a super-linear attachment rule with $\nu = 2$. Dashed lines correspond to the scaling given by Eq.~\eqref{eq:Lambda_rDMP}.}
	\label{fig:HASHIMOTO}
\end{figure}

We analyzed the spectral properties computed for the whole network, including the original core plus the bridging nodes. For the sake of comparing different amounts of bridging nodes, only the original core components of the PEV were used to compute the IPR. The LEV does not depend on this constraint. For the case $\gamma=2.3$, where a maximum $k$-core rules the PEV of both the adjacency (data not shown) and  {non-backtracking} matrices (Fig.~\ref{fig:HASHIMOTO}(a,b)) of the original core network, neither the LEV nor the IPR are significantly altered by adding a finite fraction of bridges. These results are in agreement with the role of bridging nodes in the epidemic threshold and localization of the NAV observed in the simulations of the SIS dynamics shown in Figs.~\ref{fig:compara_fs}(a,c).

\begin{figure}[hbt]
	\centering
	\includegraphics[width=0.99\linewidth]{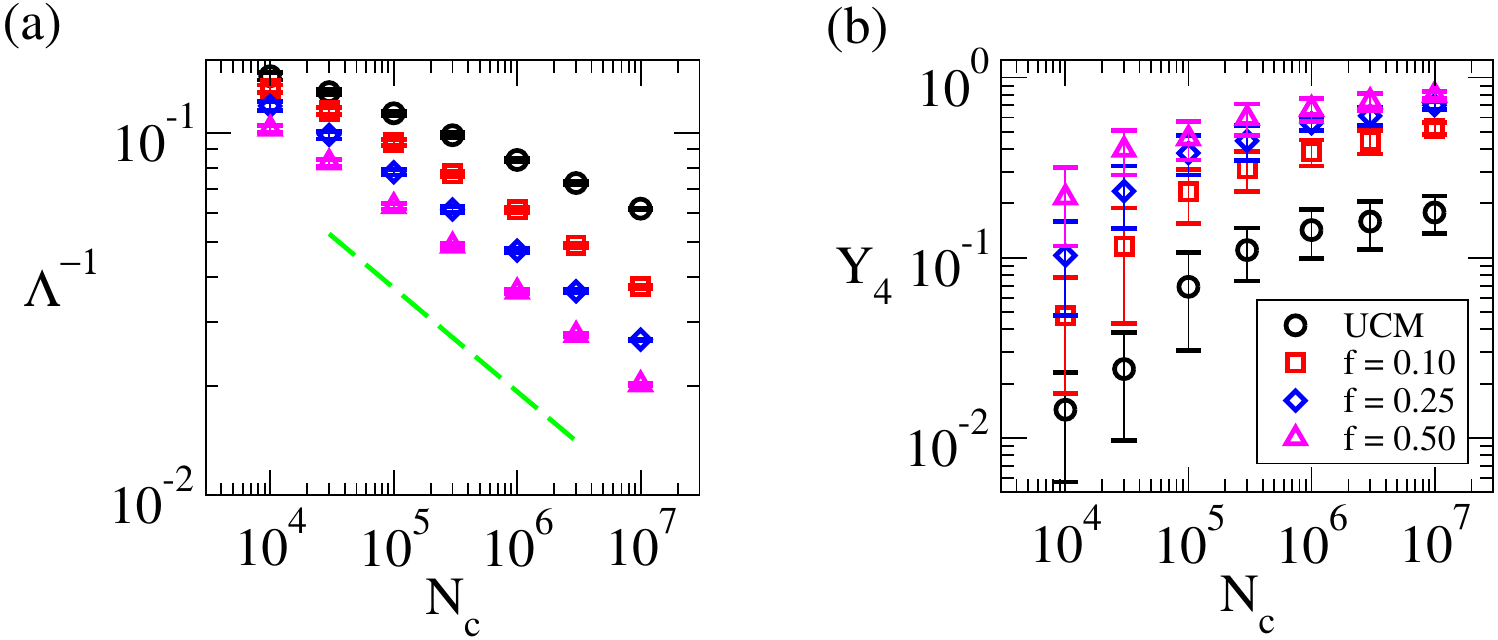}
	\caption{Spectral analysis of adjacency matrices for different amounts of bridging nodes, considering power-law networks with degree exponent $\gamma=3.5$ and a rigid cutoff for the original core. The inverse of the LEV (a) $\Lambda^{-1}$ and (b) the IPR of the PEV $Y_4$ as functions of the core size $\Ncore$ are shown for a super-linear attachment rule with $\nu = 2$. Dashed line in (a) represents the power-law given by $\Lambda_1^{-1}\sim N^{-2/7}$.}
	\label{fig:spectralAij}
\end{figure}

The spectral properties of the adjacency matrix for a degree exponent $\gamma=3.5$ are shown in Fig.~\ref{fig:spectralAij}. In this regime, the spectral properties of the original core are dominated by a localized contribution from the most connected node of the network and its neighborhood~\cite{Castellano2012,Castellano2017}. We observe that localization, measured via the IPR, increases while the inverse of the LEV decreases as bridges are added, implying that the main effect of adding bridges is to increase the localization of the PEV of the most connected node. These outcomes are qualitatively consistent with the SIS critical dynamics shown in Fig.~\ref{fig:compara_fs}, with increased localization and reduced epidemic threshold as bridging nodes are attached. This regime is fully consistent with a LEV given by $\Lambda_1\sim 1/\sqrt{\ximax}\sim \Ncore^{-\nu/2\gamma}$ for a rigid cutoff $\kmax\sim \Ncore^{1/\gamma}$ and $\gamma>\nu+1$ (Eq.~\eqref{eq:ximax}).

The spectral properties of the {non-backtracking} matrices are shown in Fig.~\ref{fig:HASHIMOTO} for different fractions of bridging nodes. For $\gamma=2.8$, a subextensive localization decaying with an exponent $Y_4\sim \Ncore^{-a}$ with $a<1$ in the original core is replaced by localization on a finite set of nodes as bridges are added, with $Y_4\sim \mathcal{O}(1)$. A similar qualitative behavior is observed for $\gamma=3.5$, but the crossover to the regime of finite IPR is slower. These scalings explain qualitatively  well the behavior of the epidemic threshold and localization observed for SIRS dynamics in Fig.~\ref{fig:sirs}.

The localization of the {non-backtracking} matrix and, consequently, the SIRS dynamics under rDMP theory can be rationalized in terms of the theoretical analysis developed in Ref.~\cite{PastorSatorras2020}, where the LEV is very well approximated by
\[\Lambda_1=\max[\Lambda^\text{un}_1,\Lambda^\text{core}_1,\Lambda^\text{oh}_1]\,,\]
where $\Lambda^\text{un}_1$ is the LEV computed under the hypothesis of uncorrelated networks, $\Lambda^\text{core}_1$ is the LEV computed only for the maximum $k$-core subgraph, and $\Lambda^\text{oh}_1$ is the LEV for the subgraph containing overlapping hubs; the latter is defined as a set of $n$ nodes of degree $K$, all sharing common $K$ neighbors of degree 2. Since bridges are the most peripheral nodes according to the $k$-core decomposition, the maximum $k$-core subgraph is unchanged with the addition of bridges. Thus, $\Lambda^\text{core}_1\simeq \av{k^2}/\av{k}$ for $\gamma<3$~\cite{Castellano2012} and is not defined for $\gamma>3$ since the maximum $k$-core is absent~\cite{Dorogovtsev2006}.

The overlapping hubs are analogous to several bridging nodes connecting a pair of hubs, so that we expect that $\Lambda_1\simeq \Lambda^\text{oh}_1$ for the regime $\gamma>2.5$, which is ruled by bridges. Using the result $\Lambda^\text{oh}_1\sim \sqrt{\ximax}$ derived in Ref.~\cite{PastorSatorras2020}, Eq.~\eqref{eq:ximax}, and the adopted cutoffs in the original core $\kmax\sim \Ncore^{1/\omega}$, where $\omega = 2$ for $\gamma<3$ and $\omega = \gamma$ for $\gamma>3$, we obtain
\begin{equation}
	\Lambda_1^{-1} \sim \left\{ \begin{matrix}
		\Ncore^{-(\gamma-1)/2\omega}&, & \gamma < \nu+1  \\
		\Ncore^{-\nu/2\omega}&, &  \gamma > \nu+1
	\end{matrix} \right.\,.
	\label{eq:Lambda_rDMP}
\end{equation}
The scaling given by Eq.~\eqref{eq:Lambda_rDMP} is compared with numerical computation of the LEV in Fig.~\ref{fig:HASHIMOTO}. We see that it works well for $\gamma<3$. It is worth commenting that the scaling of $\Lambda^\text{oh}_1$ and $\Lambda^\text{core}_1$ for $\gamma=2.3$ is practically indistinguishable, since $\Lambda^\text{oh}_1 \sim \Ncore^{0.325}$ and $\Lambda^\text{core}_1\sim \av{k^2}\sim \kmax^{3-\gamma}\sim \Ncore^{0.35}$. For $\gamma=3.5$, the discrepancy is large, indicating that isolated bridging nodes connecting hubs are not the only relevant contribution for this range of sizes. Actually, the networks have correlations, but not too strong yet, as seen in Fig.~\ref{fig:NET}. We expect that the scenario where overlapping hubs are ruling the LEV will occur for sizes much larger than those investigated here.

Reference~\cite{PastorSatorras2020} also provides an expression for the LEV associated with the   {non-backtracking} matrix for uncorrelated networks, given by
\begin{equation}
	\Lambda^\text{un}_1= \frac{\sum_{ij} (k_i-1)A_{ij} (k_j-1)}{\sum_{i} k_i(k_i-1)}\,.
\end{equation}
Despite the case $\gamma=3.5$ with bridges presenting disassortative degree correlations, as seen in Fig.~\ref{fig:NET}, we analyzed this expression for different fractions of bridging nodes. We found that $\Lambda^\text{un}_1\sim\mathcal{O}(1)$, so this expression cannot reproduce the finite-size corrections before reaching the asymptotic regime as presented in Eq.~\eqref{eq:Lambda_rDMP}.

\section{Discussion}
\label{sec:conclu}

Different centrality measures have been used as fundamental indicators for understanding dynamical processes on networks~\cite{Lu2016,DeArruda2014,DeDomenico2016,Pastor-Satorras2015}. In the context of epidemic spreading, eigenvector centralities associated with Jacobian matrices~\cite{Castellano2012,PastorSatorras2020} are commonly applied to characterize the critical behavior near epidemic eradication. Degree and $k$-core centralities have been employed to explain the nature of eigenvector localization that governs epidemic activation~\cite{Kitsak2010,Castellano2012,PastorSatorras2020}. Within this framework, low-degree nodes are relevant primarily for increasing the degrees of hubs and do not affect the maximum $k$-core, since they are peripheral according to this measure. However, low-degree nodes that bridge two important nodes can mediate feedback interactions between them, and their importance may be underestimated by these centralities.

We investigate the role of adding an extensive component of bridging nodes, representing a finite fraction of the entire network, using both stochastic simulations and mean-field theories for epidemic spreading on networks with different activation mechanisms. Three activation scenarios are considered: (i) activation driven by hubs sparsely distributed across the network; (ii) activation driven by a maximum $k$-core formed by a set of densely connected hubs; and (iii) collective activation spanning an extensive portion of the network. We focus on cases in which bridging nodes are preferentially attached to high-degree nodes of the original core.

We show that the importance of bridging nodes depends on the nature of the underlying activation mechanism. For maximum $k$-core activation, an extensive fraction of bridging nodes neither changes the epidemic threshold nor affects epidemic localization near the transition, even under a superlinear attachment rule in which bridging nodes connect to hubs proportionally to the square of their degrees. This result demonstrates that $k$-core activation is highly robust to network perturbations induced by low-degree nodes.

For epidemic activation triggered by sparsely distributed hubs, the outcome depends on the specific epidemic process. In the SIS dynamics, bridging itself is {secondary}: epidemic localization increases while the threshold decreases solely because the hubs' degrees are enhanced. The activation is consistent with a star subgraph of size $\ximax$, corresponding to the average maximum degree of the final network, with an epidemic threshold $\lambda_\text{c}\sim 1/\sqrt{\ximax}$. The localization of the principal eigenvector is quantified by the inverse participation ratio $Y_4\sim \mathcal{O}(1)$, indicating that the activity remains concentrated around a few dominant hubs.

Bridging nodes play a critical role when immunity waning is incorporated into SIRS dynamics. We consider the relevant regime in which the rate of immunity loss is smaller than the recovery rate, a condition under which the activation mechanism of the SIRS process differs substantially from that of the SIS dynamics if the degree-distribution exponent satisfies $\gamma>5/2$. For $5/2<\gamma<3$, the dynamics on the network without bridges are consistent with $k$-core activation, whereas the SIS dynamics correspond to hub activation. In these regimes, the subextensive localization observed in the original networks~\cite{Silva2022} transitions to localization within a finite set, characterized by a finite inverse participation ratio (IPR). For $\gamma>3$, the contrast becomes even more pronounced: the collective activation with a finite epidemic threshold changes to strong localization around a few leading hubs. {It is worth noting that the finite-size scaling of the effective epidemic threshold of the SIS model  on random power-law networks with $\gamma>3$ was rationalized in terms of cumulative merging percolation (CMP) in Ref.~\cite{Castellano2020}, providing quantitative expressions for the finite-size scaling of the epidemic threshold that agree better with simulations than QMF theory. However, the long-range reinfection of hubs that subsides the CPM approach is no longer relevant when bridging nodes are included, since mutual reinfection is then governed by short paths of length 2 mediated by these nodes. Indeed,} the subgraph responsible for this localization consists of overlapping hubs~\cite{PastorSatorras2020}, where pairs of hubs share a large number of common degree-2 neighbors acting as bridging nodes.

Unlike in SIS dynamics, feedback interactions mediated by bridging nodes are essential to describe epidemic activation for the SIRS model. To demonstrate this, we analyzed epidemics on networks where bridging nodes of degree 2 were transformed into two nodes of degree 1, preserving both the total number of edges and the degree sequence of the network. While in SIS dynamics no relevant change is observed between networks with bridges and those with drawbridges, in SIRS dynamics the activation mechanism reverts to that observed in the original core when the bridges are split. In other words, increasing the node degree can modify the activation mechanism, but it is the feedback interaction between hubs that ultimately governs it.

Finally, our findings for epidemic models may extend to other dynamical processes on networks. A crucial example is the refractory state in neurons~\cite{Berry1998,Weistuch2021}, in which a neuron's ability to fire another action potential is reduced or completely suppressed for a period of time. This refractory period is analogous to the temporary immunity in SIRS dynamics, whereas the redundancy of connections and correlations are key features of neuronal activity~\cite{Averbeck2006,Schneidman2003}, analogous to the multiple bridging nodes investigated in this study.


\begin{acknowledgments}
	D.H.S. and F.A.R acknowledge the support given by \textit{Fundação de Amparo à Pesquisa do Estado de São Paulo} (FAPESP)-Brazil (Grants No.  20/09835-1, 2021/00369-0, and 2013/07375).
W. C. acknowledges the financial support by INCT-DigiSaúde (CNPq Grant 408775/2024-6).
S.C.F. and F.A.R  acknowledge the financial support by the \textit{Conselho Nacional de Desenvolvimento Científico e Tecnológico} (CNPq)-Brazil (Grants No. 310984/2023-8 and 308162/2023-4), INCT-NeuroComp (CNPq Grant 408389/2024-9), and FAPESP (Grant. No. 25/24366-1). S.C.F. and W. C. than \textit{Fundação de Amparo à Pesquisa do Estado de Minas Gerais} (FAPEMIG)-Brazil (Grants No. APQ-01973-24 and APQ-03079-24).
This study was financed in part by the \textit{Coordenação de Aperfeiçoamento de Pessoal de Nível Superior} (CAPES), Brazil, Finance Code 001 .
\end{acknowledgments}

\section*{Data availability}

The data that support the findings of this article are openly available~\cite{datasets}.

\appendix

\section{Stochastic simulations}
\label{app:simu}

The stochastic simulations of the recurrent SIS and SIRS dynamics running on top of these modified networks are implemented following the optimal Gillespie algorithm (OGA)~\cite{Cota2017}. In each time step, an attempt of recovery, infection, or waning of immunity is performed. The total number of infected individuals $N_{\text{inf}}$, the total number of edges emanating from them $N_{\text{SI}}$, and the total number of recovered individuals $N_{\text{rec}}$ are determined and constantly updated. The recovery process takes place with probability
\begin{equation}
	P_{\text{I$\rightarrow$ R} }=\frac{\mu\Ninf}{\mu\Ninf +\lambda\NSI+\alpha N_{\text{rec}} }\,,
	\label{eq:OGA1}
\end{equation}
in which an infected node selected at random recovers. With probability
\begin{equation}
	P_{\text{R$\rightarrow$ S}}=\frac{\alpha N_{\text{rec}}}{\mu\Ninf+\lambda\NSI+\alpha N_{\text{rec}} }\,,
	\label{eq:OGA2}
\end{equation}
a recovered node is selected at random and returns to the susceptible state. Finally, with probability
\begin{equation}
	P_{\text{S$\rightarrow$ I}}=\frac{\lambda\NSI}{\mu\Ninf+\lambda\NSI+\alpha N_{\text{rec}} }\,,
	\label{eq:OGA3}
\end{equation}
an infected node $i$ is chosen with probability proportional to its degree. A neighbor $j$ of this node is selected at random, and if it is susceptible, $j$ changes its state to infected; otherwise, the simulation moves to the next step.

The time is always incremented by
\begin{equation}
	\delta t  = \frac{-\ln u}{\mu\Ninf+\lambda\NSI+\alpha N_{\text{rec}}}\,,
	\label{eq:GA4}
\end{equation}
in which $u$ is a pseudo random number  uniformly distributed in the interval $(0,1)$. The SIS dynamics is simulated considering only infection and healing with probabilities
\begin{equation}
	P_{\text{I$\rightarrow$ S} }=\frac{\mu\Ninf}{\mu\Ninf +\lambda\NSI }\,,
	\label{eq:OGA5}
\end{equation}
and
\begin{equation}
	P_{\text{S$\rightarrow$ I}}=\frac{\lambda\NSI}{\mu\Ninf+\lambda\NSI }\,,
	\label{eq:OGA6}
\end{equation}
respectively, and time step
\begin{equation}
	\delta t  = \frac{-\ln u}{\mu\Ninf+\lambda\NSI}\,,
	\label{eq:GA7}
\end{equation}
while the implementation rules are the same of SIRS dynamics.

\bibliography{references_bridges}

\begin{thebibliography}{46}%
\makeatletter
\providecommand \@ifxundefined [1]{%
 \@ifx{#1\undefined}
}%
\providecommand \@ifnum [1]{%
 \ifnum #1\expandafter \@firstoftwo
 \else \expandafter \@secondoftwo
 \fi
}%
\providecommand \@ifx [1]{%
 \ifx #1\expandafter \@firstoftwo
 \else \expandafter \@secondoftwo
 \fi
}%
\providecommand \natexlab [1]{#1}%
\providecommand \enquote  [1]{``#1''}%
\providecommand \bibnamefont  [1]{#1}%
\providecommand \bibfnamefont [1]{#1}%
\providecommand \citenamefont [1]{#1}%
\providecommand \href@noop [0]{\@secondoftwo}%
\providecommand \href [0]{\begingroup \@sanitize@url \@href}%
\providecommand \@href[1]{\@@startlink{#1}\@@href}%
\providecommand \@@href[1]{\endgroup#1\@@endlink}%
\providecommand \@sanitize@url [0]{\catcode `\\12\catcode `\$12\catcode
  `\&12\catcode `\#12\catcode `\^12\catcode `\_12\catcode `\%12\relax}%
\providecommand \@@startlink[1]{}%
\providecommand \@@endlink[0]{}%
\providecommand \url  [0]{\begingroup\@sanitize@url \@url }%
\providecommand \@url [1]{\endgroup\@href {#1}{\urlprefix }}%
\providecommand \urlprefix  [0]{URL }%
\providecommand \Eprint [0]{\href }%
\providecommand \doibase [0]{https://doi.org/}%
\providecommand \selectlanguage [0]{\@gobble}%
\providecommand \bibinfo  [0]{\@secondoftwo}%
\providecommand \bibfield  [0]{\@secondoftwo}%
\providecommand \translation [1]{[#1]}%
\providecommand \BibitemOpen [0]{}%
\providecommand \bibitemStop [0]{}%
\providecommand \bibitemNoStop [0]{.\EOS\space}%
\providecommand \EOS [0]{\spacefactor3000\relax}%
\providecommand \BibitemShut  [1]{\csname bibitem#1\endcsname}%
\let\auto@bib@innerbib\@empty
\bibitem [{\citenamefont {Pastor-Satorras}\ \emph {et~al.}(2015)\citenamefont
  {Pastor-Satorras}, \citenamefont {Castellano}, \citenamefont {Van~Mieghem},\
  and\ \citenamefont {Vespignani}}]{Pastor-Satorras2015}%
  \BibitemOpen
  \bibfield  {author} {\bibinfo {author} {\bibfnamefont {R.}~\bibnamefont
  {Pastor-Satorras}}, \bibinfo {author} {\bibfnamefont {C.}~\bibnamefont
  {Castellano}}, \bibinfo {author} {\bibfnamefont {P.}~\bibnamefont
  {Van~Mieghem}},\ and\ \bibinfo {author} {\bibfnamefont {A.}~\bibnamefont
  {Vespignani}},\ }\bibfield  {title} {\bibinfo {title} {Epidemic processes in
  complex networks},\ }\href {https://doi.org/10.1103/RevModPhys.87.925}
  {\bibfield  {journal} {\bibinfo  {journal} {Rev. Mod. Phys.}\ }\textbf
  {\bibinfo {volume} {87}},\ \bibinfo {pages} {925} (\bibinfo {year}
  {2015})}\BibitemShut {NoStop}%
\bibitem [{\citenamefont {Pastor-Satorras}\ and\ \citenamefont
  {Vespignani}(2001)}]{Pastor-Satorras2001b}%
  \BibitemOpen
  \bibfield  {author} {\bibinfo {author} {\bibfnamefont {R.}~\bibnamefont
  {Pastor-Satorras}}\ and\ \bibinfo {author} {\bibfnamefont {A.}~\bibnamefont
  {Vespignani}},\ }\bibfield  {title} {\bibinfo {title} {Epidemic spreading in
  scale-free networks},\ }\href {https://doi.org/10.1103/physrevlett.86.3200}
  {\bibfield  {journal} {\bibinfo  {journal} {Phys. Rev. Lett.}\ }\textbf
  {\bibinfo {volume} {86}},\ \bibinfo {pages} {3200} (\bibinfo {year}
  {2001})}\BibitemShut {NoStop}%
\bibitem [{\citenamefont {Domenico}\ \emph {et~al.}(2016)\citenamefont
  {Domenico}, \citenamefont {Granell}, \citenamefont {Porter},\ and\
  \citenamefont {Arenas}}]{DeDomenico2016}%
  \BibitemOpen
  \bibfield  {author} {\bibinfo {author} {\bibfnamefont {M.~D.}\ \bibnamefont
  {Domenico}}, \bibinfo {author} {\bibfnamefont {C.}~\bibnamefont {Granell}},
  \bibinfo {author} {\bibfnamefont {M.~A.}\ \bibnamefont {Porter}},\ and\
  \bibinfo {author} {\bibfnamefont {A.}~\bibnamefont {Arenas}},\ }\bibfield
  {title} {\bibinfo {title} {The physics of spreading processes in
  multilayer networks},\ }\href {https://doi.org/10.1038/nphys3865} {\bibfield
   {journal} {\bibinfo  {journal} {Nature Physics}\ }\textbf {\bibinfo {volume}
  {12}},\ \bibinfo {pages} {901} (\bibinfo {year} {2016})}\BibitemShut
  {NoStop}%
\bibitem [{\citenamefont {Valdano}\ \emph {et~al.}(2018)\citenamefont
  {Valdano}, \citenamefont {Fiorentin}, \citenamefont {Poletto},\ and\
  \citenamefont {Colizza}}]{Valdano2017}%
  \BibitemOpen
  \bibfield  {author} {\bibinfo {author} {\bibfnamefont {E.}~\bibnamefont
  {Valdano}}, \bibinfo {author} {\bibfnamefont {M.~R.}\ \bibnamefont
  {Fiorentin}}, \bibinfo {author} {\bibfnamefont {C.}~\bibnamefont {Poletto}},\
  and\ \bibinfo {author} {\bibfnamefont {V.}~\bibnamefont {Colizza}},\
  }\bibfield  {title} {\bibinfo {title} {Epidemic threshold in continuous-time
  evolving networks},\ }\href {https://doi.org/10.1103/PhysRevLett.120.068302}
  {\bibfield  {journal} {\bibinfo  {journal} {Physical Review Letters}\
  }\textbf {\bibinfo {volume} {120}},\ \bibinfo {pages} {068302} (\bibinfo
  {year} {2018})}\BibitemShut {NoStop}%
\bibitem [{\citenamefont {Wang}\ \emph {et~al.}(2024)\citenamefont {Wang},
  \citenamefont {Nie}, \citenamefont {Li}, \citenamefont {Lin}, \citenamefont
  {Shang}, \citenamefont {Su}, \citenamefont {Tang}, \citenamefont {Zhang},\
  and\ \citenamefont {Sun}}]{Wang2024}%
  \BibitemOpen
  \bibfield  {author} {\bibinfo {author} {\bibfnamefont {W.}~\bibnamefont
  {Wang}}, \bibinfo {author} {\bibfnamefont {Y.}~\bibnamefont {Nie}}, \bibinfo
  {author} {\bibfnamefont {W.}~\bibnamefont {Li}}, \bibinfo {author}
  {\bibfnamefont {T.}~\bibnamefont {Lin}}, \bibinfo {author} {\bibfnamefont
  {M.-S.}\ \bibnamefont {Shang}}, \bibinfo {author} {\bibfnamefont
  {S.}~\bibnamefont {Su}}, \bibinfo {author} {\bibfnamefont {Y.}~\bibnamefont
  {Tang}}, \bibinfo {author} {\bibfnamefont {Y.-C.}\ \bibnamefont {Zhang}},\
  and\ \bibinfo {author} {\bibfnamefont {G.-Q.}\ \bibnamefont {Sun}},\
  }\bibfield  {title} {\bibinfo {title} {Epidemic spreading on higher-order
  networks},\ }\href {https://doi.org/10.1016/j.physrep.2024.01.003} {\bibfield
   {journal} {\bibinfo  {journal} {Physics Reports}\ }\textbf {\bibinfo
  {volume} {1056}},\ \bibinfo {pages} {1} (\bibinfo {year} {2024})}\BibitemShut
  {NoStop}%
\bibitem [{\citenamefont {Goltsev}\ \emph {et~al.}(2012)\citenamefont
  {Goltsev}, \citenamefont {Dorogovtsev}, \citenamefont {Oliveira},\ and\
  \citenamefont {Mendes}}]{Goltsev2012}%
  \BibitemOpen
  \bibfield  {author} {\bibinfo {author} {\bibfnamefont {A.~V.}\ \bibnamefont
  {Goltsev}}, \bibinfo {author} {\bibfnamefont {S.~N.}\ \bibnamefont
  {Dorogovtsev}}, \bibinfo {author} {\bibfnamefont {J.~G.}\ \bibnamefont
  {Oliveira}},\ and\ \bibinfo {author} {\bibfnamefont {J.~F.~F.}\ \bibnamefont
  {Mendes}},\ }\bibfield  {title} {\bibinfo {title} {Localization and spreading
  of diseases in complex networks},\ }\href
  {https://doi.org/10.1103/PhysRevLett.109.128702} {\bibfield  {journal}
  {\bibinfo  {journal} {Phys. Rev. Lett.}\ }\textbf {\bibinfo {volume} {109}},\
  \bibinfo {pages} {128702} (\bibinfo {year} {2012})}\BibitemShut {NoStop}%
\bibitem [{\citenamefont {Silva}\ and\ \citenamefont
  {Ferreira}(2021)}]{Silva2021}%
  \BibitemOpen
  \bibfield  {author} {\bibinfo {author} {\bibfnamefont {D.~H.}\ \bibnamefont
  {Silva}}\ and\ \bibinfo {author} {\bibfnamefont {S.~C.}\ \bibnamefont
  {Ferreira}},\ }\bibfield  {title} {\bibinfo {title} {Dissecting localization
  phenomena of dynamical processes on networks},\ }\href
  {https://doi.org/10.1088/2632-072x/abdd98} {\bibfield  {journal} {\bibinfo
  {journal} {Journal of Physics: Complexity}\ }\textbf {\bibinfo {volume}
  {2}},\ \bibinfo {pages} {025011} (\bibinfo {year} {2021})}\BibitemShut
  {NoStop}%
\bibitem [{\citenamefont {St-Onge}\ \emph {et~al.}(2021)\citenamefont
  {St-Onge}, \citenamefont {Thibeault}, \citenamefont {Allard}, \citenamefont
  {Dub\'e},\ and\ \citenamefont {H\'ebert-Dufresne}}]{St-Onge2020a}%
  \BibitemOpen
  \bibfield  {author} {\bibinfo {author} {\bibfnamefont {G.}~\bibnamefont
  {St-Onge}}, \bibinfo {author} {\bibfnamefont {V.}~\bibnamefont {Thibeault}},
  \bibinfo {author} {\bibfnamefont {A.}~\bibnamefont {Allard}}, \bibinfo
  {author} {\bibfnamefont {L.~J.}\ \bibnamefont {Dub\'e}},\ and\ \bibinfo
  {author} {\bibfnamefont {L.}~\bibnamefont {H\'ebert-Dufresne}},\ }\bibfield
  {title} {\bibinfo {title} {Social confinement and mesoscopic localization of
  epidemics on networks},\ }\href
  {https://doi.org/10.1103/PhysRevLett.126.098301} {\bibfield  {journal}
  {\bibinfo  {journal} {Phys. Rev. Lett.}\ }\textbf {\bibinfo {volume} {126}},\
  \bibinfo {pages} {098301} (\bibinfo {year} {2021})}\BibitemShut {NoStop}%
\bibitem [{\citenamefont {Liu}\ and\ \citenamefont {Mieghem}(2019)}]{Liu2019}%
  \BibitemOpen
  \bibfield  {author} {\bibinfo {author} {\bibfnamefont {Q.}~\bibnamefont
  {Liu}}\ and\ \bibinfo {author} {\bibfnamefont {P.~V.}\ \bibnamefont
  {Mieghem}},\ }\bibfield  {title} {\bibinfo {title} {Network localization is
  unalterable by infections in bursts},\ }\href
  {https://doi.org/10.1109/TNSE.2018.2889539} {\bibfield  {journal} {\bibinfo
  {journal} {IEEE Transactions on Network Science and Engineering}\ }\textbf
  {\bibinfo {volume} {6}},\ \bibinfo {pages} {983} (\bibinfo {year}
  {2019})}\BibitemShut {NoStop}%
\bibitem [{\citenamefont {Hindes}\ and\ \citenamefont
  {Schwartz}(2016)}]{Hindes2016}%
  \BibitemOpen
  \bibfield  {author} {\bibinfo {author} {\bibfnamefont {J.}~\bibnamefont
  {Hindes}}\ and\ \bibinfo {author} {\bibfnamefont {I.~B.}\ \bibnamefont
  {Schwartz}},\ }\bibfield  {title} {\bibinfo {title} {Epidemic extinction and
  control in heterogeneous networks},\ }\href
  {https://doi.org/10.1103/PhysRevLett.117.028302} {\bibfield  {journal}
  {\bibinfo  {journal} {Physical Review Letters}\ }\textbf {\bibinfo {volume}
  {117}},\ \bibinfo {pages} {028302} (\bibinfo {year} {2016})}\BibitemShut
  {NoStop}%
\bibitem [{\citenamefont {Chatterjee}\ and\ \citenamefont
  {Durrett}(2009)}]{Chatterjee2009}%
  \BibitemOpen
  \bibfield  {author} {\bibinfo {author} {\bibfnamefont {S.}~\bibnamefont
  {Chatterjee}}\ and\ \bibinfo {author} {\bibfnamefont {R.}~\bibnamefont
  {Durrett}},\ }\bibfield  {title} {\bibinfo {title} {Contact processes on
  random graphs with power law degree distributions have critical value 0},\
  }\href {https://doi.org/10.1214/09-AOP471} {\bibfield  {journal} {\bibinfo
  {journal} {The Annals of Probability}\ }\textbf {\bibinfo {volume} {37}},\
  \bibinfo {pages} {2332 } (\bibinfo {year} {2009})}\BibitemShut {NoStop}%
\bibitem [{\citenamefont {Bogu\~n\'a}\ \emph {et~al.}(2013)\citenamefont
  {Bogu\~n\'a}, \citenamefont {Castellano},\ and\ \citenamefont
  {Pastor-Satorras}}]{Boguna2013}%
  \BibitemOpen
  \bibfield  {author} {\bibinfo {author} {\bibfnamefont {M.}~\bibnamefont
  {Bogu\~n\'a}}, \bibinfo {author} {\bibfnamefont {C.}~\bibnamefont
  {Castellano}},\ and\ \bibinfo {author} {\bibfnamefont {R.}~\bibnamefont
  {Pastor-Satorras}},\ }\bibfield  {title} {\bibinfo {title} {Nature of the
  epidemic threshold for the susceptible-infected-susceptible dynamics in
  networks},\ }\href {https://doi.org/10.1103/PhysRevLett.111.068701}
  {\bibfield  {journal} {\bibinfo  {journal} {Phys. Rev. Lett.}\ }\textbf
  {\bibinfo {volume} {111}},\ \bibinfo {pages} {068701} (\bibinfo {year}
  {2013})}\BibitemShut {NoStop}%
\bibitem [{\citenamefont {Sander}\ \emph {et~al.}(2016)\citenamefont {Sander},
  \citenamefont {Costa},\ and\ \citenamefont {Ferreira}}]{Sander2016}%
  \BibitemOpen
  \bibfield  {author} {\bibinfo {author} {\bibfnamefont {R.~S.}\ \bibnamefont
  {Sander}}, \bibinfo {author} {\bibfnamefont {G.~S.}\ \bibnamefont {Costa}},\
  and\ \bibinfo {author} {\bibfnamefont {S.~C.}\ \bibnamefont {Ferreira}},\
  }\bibfield  {title} {\bibinfo {title} {Sampling methods for the
  quasistationary regime of epidemic processes on regular and complex
  networks},\ }\href {https://doi.org/10.1103/PhysRevE.94.042308} {\bibfield
  {journal} {\bibinfo  {journal} {Phys. Rev. E}\ }\textbf {\bibinfo {volume}
  {94}},\ \bibinfo {pages} {042308} (\bibinfo {year} {2016})}\BibitemShut
  {NoStop}%
\bibitem [{\citenamefont {Lü}\ \emph {et~al.}(2016)\citenamefont {Lü},
  \citenamefont {Chen}, \citenamefont {Ren}, \citenamefont {Zhang},
  \citenamefont {Zhang},\ and\ \citenamefont {Zhou}}]{Lu2016}%
  \BibitemOpen
  \bibfield  {author} {\bibinfo {author} {\bibfnamefont {L.}~\bibnamefont
  {Lü}}, \bibinfo {author} {\bibfnamefont {D.}~\bibnamefont {Chen}}, \bibinfo
  {author} {\bibfnamefont {X.-L.}\ \bibnamefont {Ren}}, \bibinfo {author}
  {\bibfnamefont {Q.-M.}\ \bibnamefont {Zhang}}, \bibinfo {author}
  {\bibfnamefont {Y.-C.}\ \bibnamefont {Zhang}},\ and\ \bibinfo {author}
  {\bibfnamefont {T.}~\bibnamefont {Zhou}},\ }\bibfield  {title} {\bibinfo
  {title} {Vital nodes identification in complex networks},\ }\href
  {https://doi.org/10.1016/j.physrep.2016.06.007} {\bibfield  {journal}
  {\bibinfo  {journal} {Physics Reports}\ }\textbf {\bibinfo {volume} {650}},\
  \bibinfo {pages} {1} (\bibinfo {year} {2016})}\BibitemShut {NoStop}%
\bibitem [{\citenamefont {de~Arruda}\ \emph {et~al.}(2018)\citenamefont
  {de~Arruda}, \citenamefont {Rodrigues},\ and\ \citenamefont
  {Moreno}}]{DeArruda2018}%
  \BibitemOpen
  \bibfield  {author} {\bibinfo {author} {\bibfnamefont {G.~F.}\ \bibnamefont
  {de~Arruda}}, \bibinfo {author} {\bibfnamefont {F.~A.}\ \bibnamefont
  {Rodrigues}},\ and\ \bibinfo {author} {\bibfnamefont {Y.}~\bibnamefont
  {Moreno}},\ }\bibfield  {title} {\bibinfo {title} {Fundamentals of spreading
  processes in single and multilayer complex networks},\ }\href
  {https://doi.org/10.1016/j.physrep.2018.06.007} {\bibfield  {journal}
  {\bibinfo  {journal} {Physics Reports}\ }\textbf {\bibinfo {volume} {756}},\
  \bibinfo {pages} {1} (\bibinfo {year} {2018})}\BibitemShut {NoStop}%
\bibitem [{\citenamefont {Kitsak}\ \emph {et~al.}(2010)\citenamefont {Kitsak},
  \citenamefont {Havlin}, \citenamefont {Liljeros}, \citenamefont {Muchnik},
  \citenamefont {Stanley},\ and\ \citenamefont {Makse}}]{Kitsak2010}%
  \BibitemOpen
  \bibfield  {author} {\bibinfo {author} {\bibfnamefont {L.~K.}\ \bibnamefont
  {Kitsak}, \bibfnamefont {Maksimand~Gallos}}, \bibinfo {author} {\bibfnamefont
  {S.}~\bibnamefont {Havlin}}, \bibinfo {author} {\bibfnamefont
  {F.}~\bibnamefont {Liljeros}}, \bibinfo {author} {\bibfnamefont
  {L.}~\bibnamefont {Muchnik}}, \bibinfo {author} {\bibfnamefont {H.~E.}\
  \bibnamefont {Stanley}},\ and\ \bibinfo {author} {\bibfnamefont {H.~A.}\
  \bibnamefont {Makse}},\ }\bibfield  {title} {\bibinfo {title} {Identification
  of influential spreaders in complex networks},\ }\href
  {https://doi.org/10.1038/nphys1746} {\bibfield  {journal} {\bibinfo
  {journal} {Nature Physics}\ }\textbf {\bibinfo {volume} {6}},\ \bibinfo
  {pages} {888} (\bibinfo {year} {2010})}\BibitemShut {NoStop}%
\bibitem [{\citenamefont {Arruda}\ \emph {et~al.}(2014)\citenamefont {Arruda},
  \citenamefont {Barbieri}, \citenamefont {Rodríguez}, \citenamefont
  {Rodrigues}, \citenamefont {Moreno},\ and\ \citenamefont
  {Costa}}]{DeArruda2014}%
  \BibitemOpen
  \bibfield  {author} {\bibinfo {author} {\bibfnamefont {G.~F.~D.}\
  \bibnamefont {Arruda}}, \bibinfo {author} {\bibfnamefont {A.~L.}\
  \bibnamefont {Barbieri}}, \bibinfo {author} {\bibfnamefont {P.~M.}\
  \bibnamefont {Rodríguez}}, \bibinfo {author} {\bibfnamefont {F.~A.}\
  \bibnamefont {Rodrigues}}, \bibinfo {author} {\bibfnamefont {Y.}~\bibnamefont
  {Moreno}},\ and\ \bibinfo {author} {\bibfnamefont {L.~D.~F.}\ \bibnamefont
  {Costa}},\ }\bibfield  {title} {\bibinfo {title} {Role of centrality for the
  identification of influential spreaders in complex networks},\ }\bibfield
  {journal} {\bibinfo  {journal} {Physical Review E - Statistical, Nonlinear,
  and Soft Matter Physics}\ }\textbf {\bibinfo {volume} {90}},\ \href
  {https://doi.org/10.1103/PhysRevE.90.032812} {10.1103/PhysRevE.90.032812}
  (\bibinfo {year} {2014})\BibitemShut {NoStop}%
\bibitem [{\citenamefont {Dorogovtsev}\ \emph {et~al.}(2006)\citenamefont
  {Dorogovtsev}, \citenamefont {Goltsev},\ and\ \citenamefont
  {Mendes}}]{Dorogovtsev2006}%
  \BibitemOpen
  \bibfield  {author} {\bibinfo {author} {\bibfnamefont {S.~N.}\ \bibnamefont
  {Dorogovtsev}}, \bibinfo {author} {\bibfnamefont {A.~V.}\ \bibnamefont
  {Goltsev}},\ and\ \bibinfo {author} {\bibfnamefont {J.~F.~F.}\ \bibnamefont
  {Mendes}},\ }\bibfield  {title} {\bibinfo {title} {$k$-core organization of
  complex networks},\ }\href {https://doi.org/10.1103/PhysRevLett.96.040601}
  {\bibfield  {journal} {\bibinfo  {journal} {Phys. Rev. Lett.}\ }\textbf
  {\bibinfo {volume} {96}},\ \bibinfo {pages} {040601} (\bibinfo {year}
  {2006})}\BibitemShut {NoStop}%
\bibitem [{\citenamefont {Castellano}\ and\ \citenamefont
  {Pastor-Satorras}(2012)}]{Castellano2012}%
  \BibitemOpen
  \bibfield  {author} {\bibinfo {author} {\bibfnamefont {C.}~\bibnamefont
  {Castellano}}\ and\ \bibinfo {author} {\bibfnamefont {R.}~\bibnamefont
  {Pastor-Satorras}},\ }\bibfield  {title} {\bibinfo {title} {Competing
  activation mechanisms in epidemics on networks},\ }\href
  {https://doi.org/10.1038/srep00371} {\bibfield  {journal} {\bibinfo
  {journal} {Scientific Reports}\ }\textbf {\bibinfo {volume} {2}},\ \bibinfo
  {pages} {371} (\bibinfo {year} {2012})}\BibitemShut {NoStop}%
\bibitem [{\citenamefont {Cota}\ \emph {et~al.}(2018)\citenamefont {Cota},
  \citenamefont {Mata},\ and\ \citenamefont {Ferreira}}]{Cota2018}%
  \BibitemOpen
  \bibfield  {author} {\bibinfo {author} {\bibfnamefont {W.}~\bibnamefont
  {Cota}}, \bibinfo {author} {\bibfnamefont {A.~S.}\ \bibnamefont {Mata}},\
  and\ \bibinfo {author} {\bibfnamefont {S.~C.}\ \bibnamefont {Ferreira}},\
  }\bibfield  {title} {\bibinfo {title} {Robustness and fragility of the
  susceptible-infected-susceptible epidemic models on complex networks},\
  }\href {https://doi.org/10.1103/physreve.98.012310} {\bibfield  {journal}
  {\bibinfo  {journal} {Physical Review E}\ }\textbf {\bibinfo {volume} {98}},\
  \bibinfo {pages} {012310} (\bibinfo {year} {2018})}\BibitemShut {NoStop}%
\bibitem [{\citenamefont {Martin}\ \emph {et~al.}(2014)\citenamefont {Martin},
  \citenamefont {Zhang},\ and\ \citenamefont {Newman}}]{Martin2014}%
  \BibitemOpen
  \bibfield  {author} {\bibinfo {author} {\bibfnamefont {T.}~\bibnamefont
  {Martin}}, \bibinfo {author} {\bibfnamefont {X.}~\bibnamefont {Zhang}},\ and\
  \bibinfo {author} {\bibfnamefont {M.~E.~J.}\ \bibnamefont {Newman}},\
  }\bibfield  {title} {\bibinfo {title} {Localization and centrality in
  networks},\ }\href {https://doi.org/10.1103/PhysRevE.90.052808} {\bibfield
  {journal} {\bibinfo  {journal} {Physical Review E}\ }\textbf {\bibinfo
  {volume} {90}},\ \bibinfo {pages} {052808} (\bibinfo {year}
  {2014})}\BibitemShut {NoStop}%
\bibitem [{\citenamefont {Castellano}\ and\ \citenamefont
  {Pastor-Satorras}(2017)}]{Castellano2017}%
  \BibitemOpen
  \bibfield  {author} {\bibinfo {author} {\bibfnamefont {C.}~\bibnamefont
  {Castellano}}\ and\ \bibinfo {author} {\bibfnamefont {R.}~\bibnamefont
  {Pastor-Satorras}},\ }\bibfield  {title} {\bibinfo {title} {Relating
  topological determinants of complex networks to their spectral properties:
  Structural and dynamical effects},\ }\href
  {https://doi.org/10.1103/PhysRevX.7.041024} {\bibfield  {journal} {\bibinfo
  {journal} {Phys. Rev. X}\ }\textbf {\bibinfo {volume} {7}},\ \bibinfo {pages}
  {041024} (\bibinfo {year} {2017})}\BibitemShut {NoStop}%
\bibitem [{\citenamefont {Silva}\ \emph {et~al.}(2019)\citenamefont {Silva},
  \citenamefont {Ferreira}, \citenamefont {Cota}, \citenamefont
  {Pastor-Satorras},\ and\ \citenamefont {Castellano}}]{Silva2019}%
  \BibitemOpen
  \bibfield  {author} {\bibinfo {author} {\bibfnamefont {D.~H.}\ \bibnamefont
  {Silva}}, \bibinfo {author} {\bibfnamefont {S.~C.}\ \bibnamefont {Ferreira}},
  \bibinfo {author} {\bibfnamefont {W.}~\bibnamefont {Cota}}, \bibinfo {author}
  {\bibfnamefont {R.}~\bibnamefont {Pastor-Satorras}},\ and\ \bibinfo {author}
  {\bibfnamefont {C.}~\bibnamefont {Castellano}},\ }\bibfield  {title}
  {\bibinfo {title} {Spectral properties and the accuracy of mean-field
  approaches for epidemics on correlated power-law networks},\ }\href
  {https://doi.org/10.1103/PhysRevResearch.1.033024} {\bibfield  {journal}
  {\bibinfo  {journal} {Phys. Rev. Res.}\ }\textbf {\bibinfo {volume} {1}},\
  \bibinfo {pages} {033024} (\bibinfo {year} {2019})}\BibitemShut {NoStop}%
\bibitem [{\citenamefont {Pastor-Satorras}\ and\ \citenamefont
  {Castellano}(2020)}]{PastorSatorras2020}%
  \BibitemOpen
  \bibfield  {author} {\bibinfo {author} {\bibfnamefont {R.}~\bibnamefont
  {Pastor-Satorras}}\ and\ \bibinfo {author} {\bibfnamefont {C.}~\bibnamefont
  {Castellano}},\ }\bibfield  {title} {\bibinfo {title} {The localization of
  non-backtracking centrality in networks and its physical consequences},\
  }\href {https://doi.org/10.1038/s41598-020-78582-x} {\bibfield  {journal}
  {\bibinfo  {journal} {Scientific Reports}\ }\textbf {\bibinfo {volume}
  {10}},\ \bibinfo {pages} {21639} (\bibinfo {year} {2020})}\BibitemShut
  {NoStop}%
\bibitem [{\citenamefont {Ferreira}\ \emph {et~al.}(2012)\citenamefont
  {Ferreira}, \citenamefont {Castellano},\ and\ \citenamefont
  {Pastor-Satorras}}]{Ferreira2012}%
  \BibitemOpen
  \bibfield  {author} {\bibinfo {author} {\bibfnamefont {S.~C.}\ \bibnamefont
  {Ferreira}}, \bibinfo {author} {\bibfnamefont {C.}~\bibnamefont
  {Castellano}},\ and\ \bibinfo {author} {\bibfnamefont {R.}~\bibnamefont
  {Pastor-Satorras}},\ }\bibfield  {title} {\bibinfo {title} {Epidemic
  thresholds of the susceptible-infected-susceptible model on networks: A
  comparison of numerical and theoretical results},\ }\href
  {https://doi.org/10.1103/PhysRevE.86.041125} {\bibfield  {journal} {\bibinfo
  {journal} {Phys. Rev. E}\ }\textbf {\bibinfo {volume} {86}},\ \bibinfo
  {pages} {041125} (\bibinfo {year} {2012})}\BibitemShut {NoStop}%
\bibitem [{\citenamefont {Ferreira}\ \emph {et~al.}(2016)\citenamefont
  {Ferreira}, \citenamefont {Sander},\ and\ \citenamefont
  {Pastor-Satorras}}]{Ferreira2016}%
  \BibitemOpen
  \bibfield  {author} {\bibinfo {author} {\bibfnamefont {S.~C.}\ \bibnamefont
  {Ferreira}}, \bibinfo {author} {\bibfnamefont {R.~S.}\ \bibnamefont
  {Sander}},\ and\ \bibinfo {author} {\bibfnamefont {R.}~\bibnamefont
  {Pastor-Satorras}},\ }\bibfield  {title} {\bibinfo {title} {Collective versus
  hub activation of epidemic phases on networks},\ }\href
  {https://doi.org/10.1103/PhysRevE.93.032314} {\bibfield  {journal} {\bibinfo
  {journal} {Phys. Rev. E}\ }\textbf {\bibinfo {volume} {93}},\ \bibinfo
  {pages} {032314} (\bibinfo {year} {2016})}\BibitemShut {NoStop}%
\bibitem [{\citenamefont {Castellano}\ and\ \citenamefont
  {Pastor-Satorras}(2020)}]{Castellano2020}%
  \BibitemOpen
  \bibfield  {author} {\bibinfo {author} {\bibfnamefont {C.}~\bibnamefont
  {Castellano}}\ and\ \bibinfo {author} {\bibfnamefont {R.}~\bibnamefont
  {Pastor-Satorras}},\ }\bibfield  {title} {\bibinfo {title} {Cumulative
  merging percolation and the epidemic transition of the
  susceptible-infected-susceptible model in networks},\ }\href
  {https://doi.org/10.1103/PhysRevX.10.011070} {\bibfield  {journal} {\bibinfo
  {journal} {Physical Review X}\ }\textbf {\bibinfo {volume} {10}},\ \bibinfo
  {pages} {011070} (\bibinfo {year} {2020})}\BibitemShut {NoStop}%
\bibitem [{\citenamefont {Silva}\ \emph {et~al.}(2022)\citenamefont {Silva},
  \citenamefont {Silva}, \citenamefont {Rodrigues},\ and\ \citenamefont
  {Ferreira}}]{Silva2022}%
  \BibitemOpen
  \bibfield  {author} {\bibinfo {author} {\bibfnamefont {J.~C.~M.}\
  \bibnamefont {Silva}}, \bibinfo {author} {\bibfnamefont {D.~H.}\ \bibnamefont
  {Silva}}, \bibinfo {author} {\bibfnamefont {F.~A.}\ \bibnamefont
  {Rodrigues}},\ and\ \bibinfo {author} {\bibfnamefont {S.~C.}\ \bibnamefont
  {Ferreira}},\ }\bibfield  {title} {\bibinfo {title} {Comparison of
  theoretical approaches for epidemic processes with waning immunity in complex
  networks},\ }\href {https://doi.org/10.1103/PhysRevE.106.034317} {\bibfield
  {journal} {\bibinfo  {journal} {Phys. Rev. E}\ }\textbf {\bibinfo {volume}
  {106}},\ \bibinfo {pages} {034317} (\bibinfo {year} {2022})}\BibitemShut
  {NoStop}%
\bibitem [{\citenamefont {Zhao}\ \emph {et~al.}(2010)\citenamefont {Zhao},
  \citenamefont {Wu},\ and\ \citenamefont {Xu}}]{Zhao2010}%
  \BibitemOpen
  \bibfield  {author} {\bibinfo {author} {\bibfnamefont {J.}~\bibnamefont
  {Zhao}}, \bibinfo {author} {\bibfnamefont {J.}~\bibnamefont {Wu}},\ and\
  \bibinfo {author} {\bibfnamefont {K.}~\bibnamefont {Xu}},\ }\bibfield
  {title} {\bibinfo {title} {Weak ties: Subtle role of information diffusion in
  online social networks},\ }\bibfield  {journal} {\bibinfo  {journal}
  {Physical Review E - Statistical, Nonlinear, and Soft Matter Physics}\
  }\textbf {\bibinfo {volume} {82}},\ \href
  {https://doi.org/10.1103/PhysRevE.82.016105} {10.1103/PhysRevE.82.016105}
  (\bibinfo {year} {2010})\BibitemShut {NoStop}%
\bibitem [{\citenamefont {Shu}\ \emph {et~al.}(2012)\citenamefont {Shu},
  \citenamefont {Tang}, \citenamefont {Gong},\ and\ \citenamefont
  {Liu}}]{Shu2012}%
  \BibitemOpen
  \bibfield  {author} {\bibinfo {author} {\bibfnamefont {P.}~\bibnamefont
  {Shu}}, \bibinfo {author} {\bibfnamefont {M.}~\bibnamefont {Tang}}, \bibinfo
  {author} {\bibfnamefont {K.}~\bibnamefont {Gong}},\ and\ \bibinfo {author}
  {\bibfnamefont {Y.}~\bibnamefont {Liu}},\ }\bibfield  {title} {\bibinfo
  {title} {Effects of weak ties on epidemic predictability on community
  networks},\ }\bibfield  {journal} {\bibinfo  {journal} {Chaos: An
  Interdisciplinary Journal of Nonlinear Science}\ }\textbf {\bibinfo {volume}
  {22}},\ \href {https://doi.org/10.1063/1.4767955} {10.1063/1.4767955}
  (\bibinfo {year} {2012})\BibitemShut {NoStop}%
\bibitem [{\citenamefont {Gallos}\ \emph {et~al.}(2012)\citenamefont {Gallos},
  \citenamefont {Makse},\ and\ \citenamefont {Sigman}}]{Gallos2012}%
  \BibitemOpen
  \bibfield  {author} {\bibinfo {author} {\bibfnamefont {L.~K.}\ \bibnamefont
  {Gallos}}, \bibinfo {author} {\bibfnamefont {H.~A.}\ \bibnamefont {Makse}},\
  and\ \bibinfo {author} {\bibfnamefont {M.}~\bibnamefont {Sigman}},\
  }\bibfield  {title} {\bibinfo {title} {A small world of weak ties provides
  optimal global integration of self-similar modules in functional brain
  networks},\ }\href {https://doi.org/10.1073/pnas.1106612109} {\bibfield
  {journal} {\bibinfo  {journal} {Proceedings of the National Academy of
  Sciences}\ }\textbf {\bibinfo {volume} {109}},\ \bibinfo {pages} {2825}
  (\bibinfo {year} {2012})}\BibitemShut {NoStop}%
\bibitem [{\citenamefont {Ferraro}\ \emph {et~al.}(2018)\citenamefont
  {Ferraro}, \citenamefont {Moreno}, \citenamefont {Min}, \citenamefont
  {Morone}, \citenamefont {Úrsula Pérez-Ramírez}, \citenamefont
  {Pérez-Cervera}, \citenamefont {Parra}, \citenamefont {Holodny},
  \citenamefont {Canals},\ and\ \citenamefont {Makse}}]{Ferraro2018}%
  \BibitemOpen
  \bibfield  {author} {\bibinfo {author} {\bibfnamefont {G.~D.}\ \bibnamefont
  {Ferraro}}, \bibinfo {author} {\bibfnamefont {A.}~\bibnamefont {Moreno}},
  \bibinfo {author} {\bibfnamefont {B.}~\bibnamefont {Min}}, \bibinfo {author}
  {\bibfnamefont {F.}~\bibnamefont {Morone}}, \bibinfo {author} {\bibnamefont
  {Úrsula Pérez-Ramírez}}, \bibinfo {author} {\bibfnamefont
  {L.}~\bibnamefont {Pérez-Cervera}}, \bibinfo {author} {\bibfnamefont
  {L.~C.}\ \bibnamefont {Parra}}, \bibinfo {author} {\bibfnamefont
  {A.}~\bibnamefont {Holodny}}, \bibinfo {author} {\bibfnamefont
  {S.}~\bibnamefont {Canals}},\ and\ \bibinfo {author} {\bibfnamefont {H.~A.}\
  \bibnamefont {Makse}},\ }\bibfield  {title} {\bibinfo {title} {Finding
  influential nodes for integration in brain networks using optimal percolation
  theory},\ }\bibfield  {journal} {\bibinfo  {journal} {Nature Communications}\
  }\textbf {\bibinfo {volume} {9}},\ \href
  {https://doi.org/10.1038/s41467-018-04718-3} {10.1038/s41467-018-04718-3}
  (\bibinfo {year} {2018})\BibitemShut {NoStop}%
\bibitem [{\citenamefont {Serafino}\ \emph {et~al.}(2022)\citenamefont
  {Serafino}, \citenamefont {Monteiro}, \citenamefont {Luo}, \citenamefont
  {Reis}, \citenamefont {Igual}, \citenamefont {Neto}, \citenamefont
  {Travizano}, \citenamefont {Andrade},\ and\ \citenamefont
  {Makse}}]{Serafino2022}%
  \BibitemOpen
  \bibfield  {author} {\bibinfo {author} {\bibfnamefont {M.}~\bibnamefont
  {Serafino}}, \bibinfo {author} {\bibfnamefont {H.~S.}\ \bibnamefont
  {Monteiro}}, \bibinfo {author} {\bibfnamefont {S.}~\bibnamefont {Luo}},
  \bibinfo {author} {\bibfnamefont {S.~D.~S.}\ \bibnamefont {Reis}}, \bibinfo
  {author} {\bibfnamefont {C.}~\bibnamefont {Igual}}, \bibinfo {author}
  {\bibfnamefont {A.~S.~L.}\ \bibnamefont {Neto}}, \bibinfo {author}
  {\bibfnamefont {M.}~\bibnamefont {Travizano}}, \bibinfo {author}
  {\bibfnamefont {J.~S.}\ \bibnamefont {Andrade}},\ and\ \bibinfo {author}
  {\bibfnamefont {H.~A.}\ \bibnamefont {Makse}},\ }\bibfield  {title} {\bibinfo
  {title} {Digital contact tracing and network theory to stop the spread of
  {COVID}-19 using big-data on human mobility geolocalization},\ }\href
  {https://doi.org/10.1371/journal.pcbi.1009865} {\bibfield  {journal}
  {\bibinfo  {journal} {PLOS Computational Biology}\ }\textbf {\bibinfo
  {volume} {18}},\ \bibinfo {pages} {e1009865} (\bibinfo {year}
  {2022})}\BibitemShut {NoStop}%
\bibitem [{\citenamefont {Friedrich}\ \emph {et~al.}(2024)\citenamefont
  {Friedrich}, \citenamefont {Göbel}, \citenamefont {Klodt}, \citenamefont
  {Krejca},\ and\ \citenamefont {Pappik}}]{Friedrich2024}%
  \BibitemOpen
  \bibfield  {author} {\bibinfo {author} {\bibfnamefont {T.}~\bibnamefont
  {Friedrich}}, \bibinfo {author} {\bibfnamefont {A.}~\bibnamefont {Göbel}},
  \bibinfo {author} {\bibfnamefont {N.}~\bibnamefont {Klodt}}, \bibinfo
  {author} {\bibfnamefont {M.~S.}\ \bibnamefont {Krejca}},\ and\ \bibinfo
  {author} {\bibfnamefont {M.}~\bibnamefont {Pappik}},\ }\bibfield  {title}
  {\bibinfo {title} {The irrelevance of influencers: Information diffusion with
  re-activation and immunity lasts exponentially long on social network
  models},\ }\href {https://doi.org/10.1609/aaai.v38i16.29687} {\bibfield
  {journal} {\bibinfo  {journal} {Proceedings of the AAAI Conference on
  Artificial Intelligence}\ }\textbf {\bibinfo {volume} {38}},\ \bibinfo
  {pages} {17389} (\bibinfo {year} {2024})}\BibitemShut {NoStop}%
\bibitem [{\citenamefont {Catanzaro}\ \emph {et~al.}(2005)\citenamefont
  {Catanzaro}, \citenamefont {Boguñá},\ and\ \citenamefont
  {Pastor-Satorras}}]{Catanzaro2005}%
  \BibitemOpen
  \bibfield  {author} {\bibinfo {author} {\bibfnamefont {M.}~\bibnamefont
  {Catanzaro}}, \bibinfo {author} {\bibfnamefont {M.}~\bibnamefont
  {Boguñá}},\ and\ \bibinfo {author} {\bibfnamefont {R.}~\bibnamefont
  {Pastor-Satorras}},\ }\bibfield  {title} {\bibinfo {title} {Generation of
  uncorrelated random scale-free networks},\ }\href
  {https://doi.org/10.1103/PhysRevE.71.027103} {\bibfield  {journal} {\bibinfo
  {journal} {Physical Review E}\ }\textbf {\bibinfo {volume} {71}},\ \bibinfo
  {pages} {027103} (\bibinfo {year} {2005})}\BibitemShut {NoStop}%
\bibitem [{\citenamefont {Barab\'asi}\ and\ \citenamefont
  {P\'osfai}(2016)}]{barabasi2016network}%
  \BibitemOpen
  \bibfield  {author} {\bibinfo {author} {\bibfnamefont {A.-L.}\ \bibnamefont
  {Barab\'asi}}\ and\ \bibinfo {author} {\bibfnamefont {M.}~\bibnamefont
  {P\'osfai}},\ }\href {https://books.google.com.br/books?id=iLtGDQAAQBAJ}
  {\emph {\bibinfo {title} {Network science}}}\ (\bibinfo  {publisher}
  {Cambridge University Press},\ \bibinfo {address} {Cambridge, UK},\ \bibinfo
  {year} {2016})\BibitemShut {NoStop}%
\bibitem [{\citenamefont {Pastor-Satorras}\ \emph {et~al.}(2001)\citenamefont
  {Pastor-Satorras}, \citenamefont {V\'azquez},\ and\ \citenamefont
  {Vespignani}}]{Pastor_Satorras2001}%
  \BibitemOpen
  \bibfield  {author} {\bibinfo {author} {\bibfnamefont {R.}~\bibnamefont
  {Pastor-Satorras}}, \bibinfo {author} {\bibfnamefont {A.}~\bibnamefont
  {V\'azquez}},\ and\ \bibinfo {author} {\bibfnamefont {A.}~\bibnamefont
  {Vespignani}},\ }\bibfield  {title} {\bibinfo {title} {Dynamical and
  correlation properties of the internet},\ }\href
  {https://doi.org/10.1103/PhysRevLett.87.258701} {\bibfield  {journal}
  {\bibinfo  {journal} {Phys. Rev. Lett.}\ }\textbf {\bibinfo {volume} {87}},\
  \bibinfo {pages} {258701} (\bibinfo {year} {2001})}\BibitemShut {NoStop}%
\bibitem [{\citenamefont {Cota}\ and\ \citenamefont
  {Ferreira}(2017)}]{Cota2017}%
  \BibitemOpen
  \bibfield  {author} {\bibinfo {author} {\bibfnamefont {W.}~\bibnamefont
  {Cota}}\ and\ \bibinfo {author} {\bibfnamefont {S.~C.}\ \bibnamefont
  {Ferreira}},\ }\bibfield  {title} {\bibinfo {title} {Optimized {G}illespie
  algorithms for the simulation of {M}arkovian epidemic processes on large and
  heterogeneous networks},\ }\href
  {https://doi.org/https://doi.org/10.1016/j.cpc.2017.06.007} {\bibfield
  {journal} {\bibinfo  {journal} {Computer Physics Communications}\ }\textbf
  {\bibinfo {volume} {219}},\ \bibinfo {pages} {303 } (\bibinfo {year}
  {2017})}\BibitemShut {NoStop}%
\bibitem [{\citenamefont {Van~Mieghem}(2012)}]{VanMieghem2012_b}%
  \BibitemOpen
  \bibfield  {author} {\bibinfo {author} {\bibfnamefont {P.}~\bibnamefont
  {Van~Mieghem}},\ }\bibfield  {title} {\bibinfo {title} {Epidemic phase
  transition of the {SIS} type in networks},\ }\href
  {https://doi.org/10.1209/0295-5075/97/48004} {\bibfield  {journal} {\bibinfo
  {journal} {Europhysics Letters}\ }\textbf {\bibinfo {volume} {97}},\ \bibinfo
  {pages} {48004} (\bibinfo {year} {2012})}\BibitemShut {NoStop}%
\bibitem [{\citenamefont {Hashimoto}(1989)}]{Hashimoto1989}%
  \BibitemOpen
  \bibfield  {author} {\bibinfo {author} {\bibfnamefont {K.-i.}\ \bibnamefont
  {Hashimoto}},\ }\bibfield  {title} {\bibinfo {title} {Zeta functions of
  finite graphs and representations of p-adic groups},\ }\href
  {https://doi.org/10.2969/aspm/01510211} {\bibfield  {journal} {\bibinfo
  {journal} {Advanced Studies in Pure Mathematics}\ }\textbf {\bibinfo {volume}
  {15}},\ \bibinfo {pages} {211} (\bibinfo {year} {1989})}\BibitemShut
  {NoStop}%
\bibitem [{\citenamefont {Shrestha}\ \emph {et~al.}(2015)\citenamefont
  {Shrestha}, \citenamefont {Scarpino},\ and\ \citenamefont
  {Moore}}]{Munik2015}%
  \BibitemOpen
  \bibfield  {author} {\bibinfo {author} {\bibfnamefont {M.}~\bibnamefont
  {Shrestha}}, \bibinfo {author} {\bibfnamefont {S.~V.}\ \bibnamefont
  {Scarpino}},\ and\ \bibinfo {author} {\bibfnamefont {C.}~\bibnamefont
  {Moore}},\ }\bibfield  {title} {\bibinfo {title} {Message-passing approach
  for recurrent-state epidemic models on networks},\ }\href
  {https://doi.org/10.1103/PhysRevE.92.022821} {\bibfield  {journal} {\bibinfo
  {journal} {Phys. Rev. E}\ }\textbf {\bibinfo {volume} {92}},\ \bibinfo
  {pages} {022821} (\bibinfo {year} {2015})}\BibitemShut {NoStop}%
\bibitem [{\citenamefont {Berry}\ and\ \citenamefont
  {Meister}(1998)}]{Berry1998}%
  \BibitemOpen
  \bibfield  {author} {\bibinfo {author} {\bibfnamefont {M.~J.}\ \bibnamefont
  {Berry}}\ and\ \bibinfo {author} {\bibfnamefont {M.}~\bibnamefont
  {Meister}},\ }\bibfield  {title} {\bibinfo {title} {Refractoriness and neural
  precision},\ }\href {https://doi.org/10.1523/JNEUROSCI.18-06-02200.1998}
  {\bibfield  {journal} {\bibinfo  {journal} {The Journal of Neuroscience}\
  }\textbf {\bibinfo {volume} {18}},\ \bibinfo {pages} {2200} (\bibinfo {year}
  {1998})}\BibitemShut {NoStop}%
\bibitem [{\citenamefont {Weistuch}\ \emph {et~al.}(2021)\citenamefont
  {Weistuch}, \citenamefont {Mujica-Parodi},\ and\ \citenamefont
  {Dill}}]{Weistuch2021}%
  \BibitemOpen
  \bibfield  {author} {\bibinfo {author} {\bibfnamefont {C.}~\bibnamefont
  {Weistuch}}, \bibinfo {author} {\bibfnamefont {L.~R.}\ \bibnamefont
  {Mujica-Parodi}},\ and\ \bibinfo {author} {\bibfnamefont {K.}~\bibnamefont
  {Dill}},\ }\bibfield  {title} {\bibinfo {title} {The refractory period
  matters: Unifying mechanisms of macroscopic brain waves},\ }\href
  {https://doi.org/10.1162/neco_a_01371} {\bibfield  {journal} {\bibinfo
  {journal} {Neural Computation}\ }\textbf {\bibinfo {volume} {33}},\ \bibinfo
  {pages} {1145} (\bibinfo {year} {2021})}\BibitemShut {NoStop}%
\bibitem [{\citenamefont {Averbeck}\ \emph {et~al.}(2006)\citenamefont
  {Averbeck}, \citenamefont {Latham},\ and\ \citenamefont
  {Pouget}}]{Averbeck2006}%
  \BibitemOpen
  \bibfield  {author} {\bibinfo {author} {\bibfnamefont {B.~B.}\ \bibnamefont
  {Averbeck}}, \bibinfo {author} {\bibfnamefont {P.~E.}\ \bibnamefont
  {Latham}},\ and\ \bibinfo {author} {\bibfnamefont {A.}~\bibnamefont
  {Pouget}},\ }\bibfield  {title} {\bibinfo {title} {Neural correlations,
  population coding and computation},\ }\href {https://doi.org/10.1038/nrn1888}
  {\bibfield  {journal} {\bibinfo  {journal} {Nature Reviews Neuroscience}\
  }\textbf {\bibinfo {volume} {7}},\ \bibinfo {pages} {358} (\bibinfo {year}
  {2006})}\BibitemShut {NoStop}%
\bibitem [{\citenamefont {Schneidman}\ \emph {et~al.}(2003)\citenamefont
  {Schneidman}, \citenamefont {Bialek},\ and\ \citenamefont
  {Berry}}]{Schneidman2003}%
  \BibitemOpen
  \bibfield  {author} {\bibinfo {author} {\bibfnamefont {E.}~\bibnamefont
  {Schneidman}}, \bibinfo {author} {\bibfnamefont {W.}~\bibnamefont {Bialek}},\
  and\ \bibinfo {author} {\bibfnamefont {M.~J.}\ \bibnamefont {Berry}},\
  }\bibfield  {title} {\bibinfo {title} {Synergy, redundancy, and independence
  in population codes},\ }\href
  {https://doi.org/10.1523/JNEUROSCI.23-37-11539.2003} {\bibfield  {journal}
  {\bibinfo  {journal} {The Journal of Neuroscience}\ }\textbf {\bibinfo
  {volume} {23}},\ \bibinfo {pages} {11539} (\bibinfo {year}
  {2003})}\BibitemShut {NoStop}%
\bibitem [{\citenamefont {Silva}\ \emph {et~al.}(2025)\citenamefont {Silva},
  \citenamefont {Silva}, \citenamefont {Cota}, \citenamefont {Rodrigues},\ and\
  \citenamefont {Ferreira}}]{datasets}%
  \BibitemOpen
  \bibfield  {author} {\bibinfo {author} {\bibfnamefont {J.~C.~M.}\
  \bibnamefont {Silva}}, \bibinfo {author} {\bibfnamefont {D.~H.}\ \bibnamefont
  {Silva}}, \bibinfo {author} {\bibfnamefont {W.}~\bibnamefont {Cota}},
  \bibinfo {author} {\bibfnamefont {F.~A.}\ \bibnamefont {Rodrigues}},\ and\
  \bibinfo {author} {\bibfnamefont {S.~C.}\ \bibnamefont {Ferreira}},\
  }\bibfield  {title} {\bibinfo {title} {Data for "impacts of bridging nodes on
  the epidemic activation mechanisms"},\ }\href
  {https://doi.org/10.5281/zenodo.17612065} {10.5281/zenodo.17612065} (\bibinfo
  {year} {2025})\BibitemShut {NoStop}%
\end{thebibliography}%

\end{document}